\documentclass[11pt,pdflatex,sn-mathphys-ay,iicol]{sn-jnl}

\usepackage{graphicx}%
\usepackage{multirow}%
\usepackage{amsmath,amssymb,amsfonts}%
\usepackage{amsthm}%
\usepackage{mathrsfs}%
\usepackage[title]{appendix}%
\usepackage[pdftex,dvipsnames]{xcolor} 
\usepackage{textcomp}%
\usepackage{manyfoot}%
\usepackage{booktabs}%
\usepackage{algorithm}%
\usepackage{algorithmicx}%
\usepackage{algpseudocode}%
\usepackage{listings}%
\usepackage{bbm}
\usepackage{multicol}
\usepackage{makecell}
\usepackage{array}
\usepackage{float}
\usepackage{geometry}
\usepackage{array}    
\usepackage{booktabs} 

\usepackage{titlesec}

\usepackage{verbatim}

\usepackage{lipsum}                     
\usepackage{xargs}                      
\usepackage[colorinlistoftodos,prependcaption,textsize=tiny]{todonotes}

\newtheorem{proposition}{Proposition}

\theoremstyle{thmstylethree}%
\newtheorem{definition}{Definition}%

\titleclass{\subsubsubsection}{straight}[\subsubsection]
\newcounter{subsubsubsection}[subsubsection]

\renewcommand\thesubsubsubsection{\thesubsubsection.\arabic{subsubsubsection}}
\titleformat{\subsubsubsection}[runin]
  {\normalfont\normalsize\bfseries}{\thesubsubsubsection}{1em}{}
\titlespacing*{\subsubsubsection}{0pt}{3.25ex plus 1ex minus .2ex}{1em}

\title{\textbf{A multiscale method for data collected from network edges via the line graph}}

\author[1,*]{\textbf{Dingjia Cao} 
\author[1]{\textbf{Marina I. Knight}}
\author[2]{\textbf{Guy P. Nason}}}

\begin{document}

\maketitle

\begin{abstract}
\noindent
    \textbf{Abstract.} 
Data collected over networks can be modelled as noisy observations of an unknown function over the nodes of a graph or network structure, fully described by its nodes and their connections, the edges. In this context, function estimation has been proposed in the literature and typically makes use of the network topology such as relative node arrangement, often using given or artificially constructed node Euclidean coordinates. However, networks that arise in fields such as hydrology (for example, river networks) present features that challenge these established modelling setups since the target function may naturally live on edges (e.g., river flow) and/or the node-oriented modelling uses noisy edge data as weights. This work tackles these challenges and develops a novel lifting scheme along with its associated (second) generation wavelets that permit data decomposition across the network edges. The transform, which we refer to under the acronym LG-LOCAAT, makes use of a line graph construction that first maps the data in the line graph domain. We thoroughly investigate the proposed algorithm's properties and illustrate its performance versus existing methodologies. We conclude with an application pertaining to hydrology that involves the denoising of a water quality index over the England river network, backed up by a simulation study for a river flow dataset.
\end{abstract}


\vfill
\noindent
\hrule 
\vspace{1em} 
\noindent
$^1$ Department of Mathematics, University of York, York, UK\\
\noindent
$^2$ Department of Mathematics, Imperial College London, UK\\
$^*$ Corresponding author: dingjia.cao@york.ac.uk\\




\section{Introduction}\label{sec1}
Networks, as collections of interconnected objects mapped onto graph structures, allow us to understand the behaviour of complex systems and are relevant for many fields amongst which climate and hydrology are the focus of this work. Access to the network architecture has the potential to capture and model irregular data structures, for example, social and economic processes \citep{jackson2002evolution}; brain networks \citep{bullmore2009complex}; epidemiological networks \citep{danon2011networks}; and traffic networks \citep{zhang2022complex}. In recent years, the statistical modelling of networks has increasingly gained momentum, with applications such as subgraph density estimation \citep{chang2022estimation}; nonparametric regression on networks \citep{severn2021non}; spatio-temporal modelling on networks \citep{knight2019generalised}.

Although the current literature is rich when data is collected from the network's vertex space, in reality, data may naturally come
from the edge set rather than the vertex set, for example, traffic flow data \citep{lakhina2004structural}; data from river networks (\citealt{cressie2006spatial}; \citealt{park2022lifting}); and fish
species distribution \citep{buisson2008modelling}.

Let us next introduce a hydrological dataset that motivates our work.

\subsection{Data collected from the network edges: a hydrological example}
Hydrological data hold the key to ultimately understanding and/or forecasting related processes and their associated changes  \citep{mcmillan2018hydrological}. In this work, we focus on the dissolved oxygen (DO) measured by the amount of oxygen in the water, which is an indicator of water quality. Understanding DO data levels in their geographical context in turn allows us to assess the influence of weather and human behaviour (e.g., seasonal changes and pollution, respectively) on river ecology, beneficial from societal and economic perspectives.

\begin{figure}
    \centering
    \includegraphics[width=6cm, height=7cm]{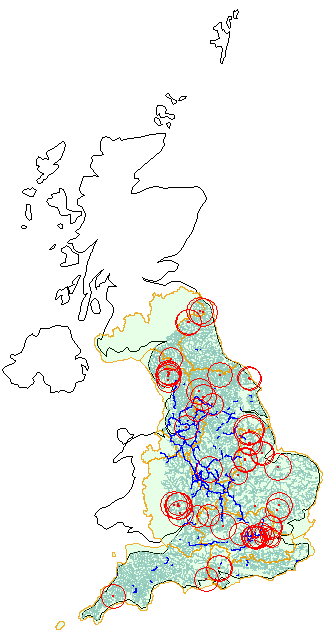}
    \caption{England river network geometry (data collected on 10/05/2024). The green-coloured areas bounded by orange curves are different river basins. The light-blue grey curves are the river water bodies, and the blue ones are canal water bodies. The red dots are the stations that collected data, and the red circle circumference indicates the associated DO data value (larger circles indicate larger DO values). }
    \label{fig:UK_river_network}
\end{figure}

Figure \ref{fig:UK_river_network} provides a visualisation of DO levels over the river network in England. The river network can be separated into 10 river basins, and the 60 sampled stations are
distributed in 9 out of 10 river basins. The size of the red circles is determined by the DO data values, where a larger circunference indicates a larger value. The dataset can be found in an open-resource website (\url{https://environment.data.gov.uk/hydrology/explore}).

Naturally, it would be desirable to consider the river system as a network structure, where the river conjunctions, sources and mouths can be considered as the vertices, and each river will be recognised as an edge. Since the data collecting stations are next to (or close to) corresponding rivers, it is more reasonable to consider the data as collected from network edges instead of vertices.

\subsection{Aim and structure of the paper}\label{subsec_intro}
As hydrological data measurements are typically noise contaminated (see \cite{mcmillan2018hydrological} on imprecise measurements and data management errors), framing statistical modelling as nonparametric regression problems is highly valuable for the understanding of hydrological processes. 

When compared with the analysis of data collected from the network vertex space, the analysis of data collected from network edges calls for new techniques, and in its turn reaches across many application fields. Wavelets, as a set of tools that allow us to capture the features of underlying functions in a multiscale fashion, are a desirable tool for performing the statistical analyses on hydrological processes. Due to the irregular nature of the network-structured data, the lifting scheme introduced by \cite{sweldens1996lifting,sweldens1998lifting} can be the key to such analyses. However, it is not trivial to apply the existing methods to edge-based data (see the heavy adaptation in \cite{park2022lifting}), due to edge topology/geometry considerations, as opposed to the much simpler case when data have been collected over one-dimensional locations, or even over the vertex spaces \citep{jansen2009multiscale}. Similarly, wavelets constructed on the vertex set of a network structure, referred to as the ‘graph wavelets’ among the signal processing community \citep{shuman2016vertex, stankovic2020vertex}, cannot be directly employed on network edges.

Our contribution is a new multiscale methodology capable of denoising corrupted data collected from the network edges. The proposed method involves a transform based on the line graph technique, followed by a variant of the lifting one coefficient at a time (LOCAAT) scheme of \cite{jansen2009multiscale} operating in the line graph space. Our proposed method allows for the construction of wavelet functions as well as for multiscale function representations, and has competitive denoising performance when compared with current literature methods that adjust existing vertex-based methodology for use on edges, such as  \cite{park2022lifting}.

The paper is organised as follows. Section~\ref{sec:litreview} gives a brief literature review of the key concepts that set the scene for introducing the proposed algorithm and its properties in Section~\ref{sec:algo} (see also Appendices~\ref{app:theory} and~\ref{app:proofs}). Section~\ref{sec:simstudy} sets up a thorough simulation study that explores aspects such as algorithm sparsity, stability and its denoising properties (Appendix~\ref{app:simresults} gives further details). Section~\ref{sec:realdata} investigates the DO dataset along with a simulated flow dataset previously introduced in the literature, and Section~\ref{sec:concl} concludes the work.


\section{Background} \label{sec:litreview}
In this section we introduce background knowledge for our methodology, ranging from graph theory to a variant of the wavelet lifting scheme. 

\subsection{Graphs and metrized graphs}
\label{sec:graph}
Mathematically, a graph $G$ can be represented as an ordered pair $G = (\mathcal{V}, \mathcal{E})$, where $\mathcal{V} = \{v_1, ..., v_n\}$ is the set of vertices and $\mathcal{E} =
\{e_1, ..., e_m\}$ is the set of edges, see e.g., \cite{bondy1976graph}. The cardinality of the vertex and the edge sets is $|\mathcal{V}| = n$ and $|\mathcal{E}| = m$, respectively. If the $k$-th edge indicates the connection between $i$-th and $j$-th vertices, then we write $e_k = \{v_i, v_j\}$, which is an unordered pair of these two vertices ($\{v_i, v_j\} = \{v_j, v_i\}$). In this case, we say $v_j$ is a neighbour of $v_i$, and vice versa. We denote the set of neighbouring vertices of the vertex $v_i$ by $\mathcal{N}^{\mathcal{V}}_i = \{v_j \, | \, v_j \in \mathcal{V}; \, \{v_i, v_j\} \in \mathcal{E}\}$.

A \textbf{weighted graph} $G^{\mathcal{\omega}}$ is modelled by an ordered pair $(G, \omega)$, where $\omega: \mathcal{E} \longrightarrow \mathbb{R}$ is a function that associates a weight to each edge \citep{bondy1976graph}. To fit our framework, we let $\omega: \mathcal{E} \longrightarrow \mathbb{R}_{+}^{\star}$ for any undirected graph and for the edge $e_k = \{v_i, v_j\} \in \mathcal{E}$, the notation $\omega_k$ or $\omega_{ij}$ will be used interchangeably to represent the weight defined on this edge. In the literature on graph theory, there are many different ways to consider the weights associated with edges; for example, weights can be the cost of the travelling salesman problem \citep{bondy1976graph}, resistance in electronic networks \citep{bollobas2013modern}, or traffic flow in a road network. The size of the weights has different (sometimes opposite) effects in various applications, e.g., a large weight on an edge indicates a weak connection between the vertices when the weights represent cost, while in the case of traffic flow, it indicates a strong link.

For the purpose of our work, we let the connection between two vertices $v_i$ and $v_j$ increase monotonically with the weight of the edge $e_k = \{v_i, v_j\}$. Therefore, if $v_i$ and $v_j$ are close to each other, then the weight $\omega_k$ tends to be large. Weighted graphs are widely used in many classical graph theory problems. However, recently, more focus has been put on metric graphs, due to their wide ranging applications, for example, for Hilbert spaces defined on graphs \citep{kuchment2003quantum} and for the analysis of electronic networks \citep{baker2006metrized}. 
In a nutshell, the metrized graph gives geometric properties for a weighted graph and allows us to define and analyse functions in the graph domain. 
A metrized graph $\Gamma$ of a weighted graph $G^{\omega} = (G, \omega)$ arises from a pair $(G, \ell)$, where $\ell: \mathcal{E} \longrightarrow \mathbb{R}_+$ is a function that assigns lengths to all edges, in the following way. Here we adopt the construction from \cite{baker2006metrized} and define the length as inverse weight, such that $\ell(e) = 1/\omega(e)$, for all $e\in \mathcal{E}$. Then to each edge $e$, we associate a line segment (or equivalently, a one-dimensional interval) of length $\ell (e)$ and identify the ends of distinct line segments if they correspond to the same vertex $v \in \mathcal{V}$. For the length of a certain edge, say the $k$-th edge $e_k$, we write $\ell_k$ for convenience. The space $\Gamma(G^{\omega})$ is the space that contains the points in any of these line segment intervals, and $G^{\omega}$ is referred to as a model for $\Gamma(G^{\omega})$ (simply $\Gamma$ if there is no ambiguity). The distance between two points in $\Gamma$ is defined as the length of the shortest path between them along the line segments traversed. This distance is referred to as the path metric, and $\Gamma$ with this metric form a metrized graph.

\cite{baker2006metrized} also introduced a way to define a vertex set and the associated edge set on $\Gamma$, summarised below. 
\begin{itemize}
    \item \textbf{Defining a vertex set on $\Gamma$:}

    Let $\text{Vex}(\Gamma)$ be any finite, non-empty subset of $\Gamma$, such that $\text{Vex}(\Gamma)$ contains all points with $n_\mathbf{p} \neq 2$,
    where $n_\mathbf{p}$ denotes the number of the directions by which a path can leave the point $\mathbf{p}$, which is referred to as the valence in \cite{baker2006metrized}. Hence, $\Gamma\backslash \text{Vex}(\Gamma)$ is a finite, disjoint union of subspaces $\{U_k\}_k$ isometric to open intervals, where $U_k$ can be considered as a neighbourhood of a point $\mathbf{p}_k \in \Gamma$.
    

    \item \textbf{Defining an edge set on $\Gamma$:} 

    Once a vertex set has been defined, the associated (metrized) edge set can be constructed based on it.
    We define the topological closure of ${U}_k$, isometric to a line segment, to be $e^{\text{met}}_k = \overline{U}_k$ as the $k$-th edge of the metrized graph. Alternatively, denote the metrized $k$-th edge $e^{\text{met}}_k$ of $e_k = \{v_i, v_j\}$ as $[v_i, v_j]$ for convenience. Additionally, any distinct metrized edges $e^{\text{met}}_{k}$ and $e^{\text{met}}_{k'}$ ($k\neq k'$) intersect in at most one point.
\end{itemize}



    For a metrized graph $\Gamma$ corresponding to the original graph $G^\omega$, the space $L^2(\Gamma)$ \citep{kuchment2003quantum} consists of functions that are measurable and integrable on each metrized edge, such that
    \begin{align}
        \lVert f \rVert^2_{L^2(\Gamma)} = \sum_{e\in \mathcal{E}} \lVert f \rVert^2_{L^2(e^{\text{met}})} < \infty,  \nonumber
    \end{align}
    where the metrized edge $e^{\text{met}}$ corresponding to the original edge $e\in \mathcal{E}$ is isometric to a line segment $[0, \ell(e^{\text{met}})]$ and $\ell(e^{\text{met}})$ is the length of this edge.

The space $L^2(\Gamma)$ can be considered as the orthogonal direct sum of $L^2(e^{\text{met}})$, see \cite{kuchment2003quantum}. Consequently, since $e^{\text{met}}$ is isometric to a closed interval on $\mathbb{R}$, the inner product and orthogonality can be defined as in $L^2(\mathbb{R})$.

\subsection{A wavelet lifting transform for graphs} \label{sec:LOCAAT}
Our algorithm will make use of the LOCAAT (\textbf{L}ifting \textbf{O}ne \textbf{C}oefficient \textbf{A}t \textbf{A} \textbf{T}ime) transform of \cite{jansen2009multiscale}, which is a variant of the lifting scheme \citep{sweldens1998lifting} that performs a split stage that isolates one point at each step, as opposed to carrying out an odds and evens split. Assuming the graph is underpinned by a metric such as path length or Euclidean distance, LOCAAT provides a wavelet-like decomposition that only relies on the network topology and on the inter-vertex distance information. 

For a function $f^{\mathcal{V}}: \mathcal{V} \rightarrow \mathbb{R}$, denote by $f^\mathcal{V}_i$ the observation at the $i$-th vertex $v_i\in \mathcal{V}$, for all $i\in \{1,...,n\}$. The initial scaling coefficients are obtained by taking $c^{\mathcal{V}}_{i, n} := f^{\mathcal{V}}_{i}$, for all $i\in \{1,...,n\}$. Then carry out the stage-$n$ LOCAAT transform, consisting of four steps as follows.

\begin{itemize}
    \item \textbf{Split:} One vertex, namely $v_{i_n}$, will be chosen to be predicted and then to be removed. 

    \item \textbf{Predict:} The observation value $f^{\mathcal{V}}_{i_n}$ associated with $v_{i_n}$ will be predicted using its neighbouring nodes ($\mathcal{N}^{\mathcal{V}}_{i_n,n}$), and the residue denoted by
    \begin{align}
    \label{equ:LOCAAT_predict}
        d^{\mathcal{V}}_{i_n} = c^{\mathcal{V}}_{i_n, n} - \sum_{j : \,v_j \in \mathcal{N}^{\mathcal{V}}_{i_n,n}} a^{\mathcal{V}}_{j, n} c^{\mathcal{V}}_{j, n},
    \end{align}
    where $\{a^{\mathcal{V}}_{j, n}\}_{j : \,v_j \in \mathcal{N}^{\mathcal{V}}_{i_n ,n}}$ is the prediction filter at stage-$n$. 
    The difference $d^{\mathcal{V}}_{i_n}$ between the observation value and the prediction value is interpreted as the detail (wavelet) coefficient.

    \item \textbf{Update:} The scaling coefficients associated with the neighbouring vertices of $v_{i_n}$ will be updated from stage-$n$ to stage-$(n-1)$ as follows
    \begin{align}
    \label{equ:LOCAAT_update}
        c^{\mathcal{V}}_{j, n-1} = c^{\mathcal{V}}_{j, n} + b^{\mathcal{V}}_{j, n} d^{\mathcal{V}}_{i_n}, \,\,\, \text{$\forall j \in \mathcal{N}^{\mathcal{V}}_{i_n ,n}$},
    \end{align}
    and the remaining scaling coefficients will not be changed. Thus, $\forall k\notin \mathcal{N}^{\mathcal{V}}_{i_n ,n}$, we let $c^{\mathcal{V}}_{k, n-1} := c^{\mathcal{V}}_{k, n}$. In addition, the neighbouring integral values will also be updated to account for the loss of vertex $v_{i_n}$, such that
    \begin{align}
        \label{equ:LOCAAT_integral_update}
        I^\mathcal{V}_{j, n-1} = I^\mathcal{V}_{j, n} + a^\mathcal{V}_{j, n} I^\mathcal{V}_{i_n, n}, \,\,\, \forall j \in \mathcal{N}^{\mathcal{V}}_{i_n ,n}.
    \end{align}
    For $\forall k\notin \mathcal{N}^{\mathcal{V}}_{i_n ,n}$, we let $I^{\mathcal{V}}_{k, n-1} := I^{\mathcal{V}}_{k, n}$.

    \item \textbf{Relink:} 
    Once the vertex $v_{i_n}$ has been removed, \cite{jansen2009multiscale} suggest performing a minimum spanning tree among the neighbouring node set $\{v_j\in \mathcal{N}^{\mathcal{V}}_{i_n ,n}\}$ for relinkage. 
    
    \item \textbf{Iterate :} Repeat the steps above until a stopping time $\tau$ set in advance. Typically, iterate the above steps $(n-2)$ times, resulting in $(n-2)$ detail coefficients and $\tau = 2$ scaling coefficients. At each stage-$r$, the split-predict-update-relink procedure is of the same form as the one presented above (stage-$n$), and we only change notation $n$ to $r$, e.g., $i_r$ indicates the vertex chosen for removal at stage-$r$.
\end{itemize}
The LOCAAT transform is invertible through iteratively undoing each of its steps, or equivalently by noticing that being a linear transform, it can be expressed using matrix multiplication \citep{nunes2006adaptive}.

\subsection{Nonparametric regression over graphs}\label{sec:nonparamregr}
In general terms, nonparametric regression problems amount to typically modellling the observations $\{(x_i, f_i)\}_{i=1}^n$ as being additively contaminated by noise $\{\epsilon_i\}_{i=1}^n$
\begin{align}
\label{nonparametric}
    f_i = g(x_i) + \epsilon_i,
\end{align}
where $x_i\in \Omega$ is the data location from a certain bounded domain $\Omega$, and $g: \Omega \longrightarrow \mathbb{R}$ is a true, unknown function that we aim to estimate. The noise is assumed to be a set of independent, identically distributed (iid) normal random variables, $\epsilon_i \stackrel{\text{iid}}{\sim} N(0, \sigma^2)$, where $\sigma^2$ is finite and usually unknown. The bounded domain can be a one-dimensional interval $\Omega=[0,1]$ \citep{nunes2006adaptive}, high-dimensional Euclidean space, or graph vertex domain $\Omega = \mathcal{V}$ for a graph $G= (\mathcal{V}, \mathcal{E})$ \citep{jansen2009multiscale, mahadevan2010multiscale}. 

The classical LOCAAT transform introduced in Section \ref{sec:LOCAAT} can be used for denoising the set of nodal observations $\underline {f}^{\mathcal{V}}=\{f_i^{\mathcal{V}} \}_{i=1}^n$, by first projecting them into the lifting domain and obtaining a set of wavelet coefficients $\underline {d}^{\mathcal{V}}$. To this end, an empirical Bayes thresholding approach adapted to the lifting domain can be taken \citep{johnstone2004needles, nunes2006adaptive}, followed by inverting LOCAAT. This leads to obtaining an estimator $\underline{\hat{g}}^{\mathcal{V}}$ for the true (unknown) function $g^{\mathcal{V}}$, where the superscript indicates the fact that the observations, and consequently the estimates, are obtained over the network vertices. 

For the problem tackled through this work, the observations are collected over the graph edges $\Omega=\mathcal{E}$, and a direct application of LOCAAT for projection in the wavelet domain is not possible.


\section{Proposed LG-LOCAAT algorithm}\label{sec:algo}

We introduce an approach to carry out multiscale analysis on data collected from the edges of a network modelled as a graph $G=(\mathcal{V},\mathcal{E})$. Denote a set of observations on a function $g^\mathcal{E}$ collected from the graph edges by
$\{g^{\mathcal{E}}_k\}_{k=1}^m$. The method we propose involves initially applying a line graph transform in order to shift emphasis from (original) edges to (new) vertices, followed by a version of the LOCAAT on the vertices in the line graph domain. Therefore, we refer to our proposed technique as \textbf{L}ine \textbf{G}raph \textbf{LOCAAT} (LG-LOCAAT).

Specifically, the line graph of a graph $G$, denoted $\mathbf{LG}(G)$, is the graph whose vertex set is bijective to the edge set $\mathcal{E}$ of $G$. For convenience, we write $\mathbf{LG}(G) = G^* = (\mathcal{V}^*, \mathcal{E}^*)$, and we have $\mathcal{V}^* \longleftrightarrow \mathcal{E}$, where `$\longleftrightarrow$' indicates a bijection between two sets. Moreover, we simply let a new vertex $v_k^*$ correspond to the $k$-th original edge $e_k \in \mathcal{E}$, for any $k\in \{1,..., |\mathcal{E}|\}$. Although the correspondence of subscripts can be any permutation, it will not affect the result of our work since LOCAAT does not rely on this ordering. The new edge set $\mathcal{E}^*$ can be defined as
\begin{equation}
    \mathcal{E}^* = \left\{ \{ v^*_k, v^*_l \} \, | \, \text{$e_k, e_l \in \mathcal{E}$ and $|e_k \cap e_l| = 1$} \right\}.  \nonumber
\end{equation}
Here the cardinality of the intersection of two (original) edges being equal to one indicates that $e_k$ and $e_l$ share exactly one common vertex in the graph $G$.
Similarly, the neighbourhood $\mathcal{N}^{\mathcal{V}^*}_k$ of a vertex $v^*_k$ in the line graph is defined as the set of new vertices which are connected with $v^*_k$ by an edge $e^*\in \mathcal{E}^*$, mathematically represented as 
\begin{align}
\label{equ:LG-neighbour1}
    \mathcal{N}^{\mathcal{V}^*}_k &:= \{v^*_{s} \in \mathcal{V}^* \, | \, \{v^*_k, v^*_s\} \in \mathcal{E}^* \}   \\
\label{equ:LG-neighbour2}
    &= \{v^*_{s} \in \mathcal{V}^* \, | \, s\neq k ; \, e_k \cap e_s \neq \emptyset; \, e_k, e_s \in \mathcal{E} \}
\end{align}

Note that line graphs reveal only combinatorial information between edges. To perform LOCAAT on a line graph, we will use the path length or Euclidean distances between the new vertices.


Before introducing our proposed algorithm, let us first take a function representation perspective. Following the notation of \cite{diestel2005graph}, suppose that we have a real-valued function $g^{\mathcal{E}} \in \mathfrak{E}$, where $\mathfrak{E}$ is the vector space that contains all functions that map $\mathcal{E}$ to $\mathbb{R}$, where $\mathcal{E}$ is the edge set of the original  graph $G$.
Similarly, a vector space $\mathfrak{V}^*$ containing all functions defined on the new vertex set of the line graph $\mathbf{LG}(G)=G^*$ can be also defined. The bijection between $\mathcal{E}$ and $\mathcal{V}^*$ indicates that the two vector spaces $\mathfrak{E}$ and $\mathfrak{V}^*$ are isomorphic, denoted as $\mathfrak{E} \approx \mathfrak{V}^*$ \citep{roman2005advanced}. Thus, there exists a function $g^{\mathcal{V}^*}\in \mathfrak{V}^*$, where $g^{\mathcal{V}^*}(v^*_k) = g^{\mathcal{E}}(e_k)$ if $v^*_k$ is the image of $e_k$ according to the line graph transform. We denote $g^{\mathcal{V}^*} \equiv g^{\mathcal{E}}$ if they satisfy this condition for all $k$. This isomorphism has many advantages for our construction. First of all, since additivity and scalar multiplication are maintained in isomorphic vector spaces \citep{roman2005advanced}, performing a linear combination of a set of vectors $\{\mathbf{u}_s \in \mathfrak{V}^*\}_{s}$ is equivalent to performing the same linear combination of a set of vectors $\{\mathbf{w}_s \in \mathfrak{E}\}_{s}$, where $\mathbf{u}_s \equiv\mathbf{w}_s$. Secondly, if $\{\mathbf{u}_s\}_{s=1}^p$ is a basis defined on $\mathfrak{V}^*$, then $\{\mathbf{w}_s\}_{s=1}^p$ is a basis defined on $\mathfrak{E}$, and $\mathbf{u}_s \equiv \mathbf{w}_s$. 
As a result, it is feasible to design a multiresolution (MRA) analysis for $g^{\mathcal{E}}$ based on $\mathcal{V}^*$. 

Incorporating the results from \cite{sweldens1996lifting,sweldens1998lifting} and the LOCAAT framework, we summarise the conditions for suitably designing primal and dual scaling functions for the vertex set $\mathcal{V}^*$ of the line graph $G^*$, as follows. The set of dual scaling functions $\{\Tilde{\varphi}^{\mathcal{V}^*}_{k,m}\}_{k=1}^m$ should satisfy $\langle \Tilde{\varphi}^{\mathcal{V}^*}_{k,m}, g^{\mathcal{V}^*} \rangle = g^{\mathcal{V}^*}_k$, where $\{g^{\mathcal{V}^*}_k\}_{k=1}^m$ is the set of original edge observations, now mapped onto the line graph vertices. 
This condition allows us to start with the observation values as the initial scaling coefficients, and then to perform the lifting scheme. The primal scaling functions $\{{\varphi}^{\mathcal{V}^*}_{k,m}\}_{k=1}^m$ are designed such that for each $k\in \{1,...,m\}$, we have ${\varphi}^{\mathcal{V}^*}_{k,m}(v^*_k) = 1$ and ${\varphi}^{\mathcal{V}^*}_{k,m}(v^*_s) = 0$ for all $s\neq k$. The set of primal and dual initial scaling functions has to satisfy the biorthogonality property, such that $\langle \Tilde{\varphi}^{\mathcal{V}^*}_{k,m}, {\varphi}^{\mathcal{V}^*}_{k',m} \rangle = \delta_{kk'}$, where $\delta_{kk'}$ is the Kronecker delta. 

For a graph $G$, 
we denote the associated metric space of its line graph by $(\mathcal{V}^*, \text{dist}_{\mathcal{V}^*})$
and its metrized version by $(\Gamma^*, \text{dist})$. 
The distance measure ($\text{dist}$) between two metrized vertices is defined as the length of the corresponding metrized edge $e^{*\text{met}}_{k}$, for any edge $e^{*}_{k} = \{v^*_i, v^*_j\}$.
We denote a point in the metrized graph space as $\mathbf{p}^* \in \Gamma^*$, and denote by $\mathbf{p}^*_{v^*_k}$ the point that corresponds to $v^*_k$. Then we can define a set of partitionings, $\{\mathbf{P}^*_s\}_{s=1}^m$, of the metrized line graph such that
\begin{align}
\label{equ:LG-partitioning}
    \mathbf{P}^*_s = \Big\{\mathbf{p}^*\in \Gamma^* \, | \, &\text{dist}(\mathbf{p}^*, \mathbf{p}^*_{v^*_s}) < \text{dist}(\mathbf{p}^*, \mathbf{p}^*_{v^*_{s'}}),  \nonumber \\
    &\,\,\text{$\forall s' \in \{1,...,m\}\backslash \{s\}$} \Big\},
\end{align}
e.g., the middle points of the metrized edges $e^{*\text{met}}$ generate a partitioning. 
In the same vein as \cite{jansen2009multiscale}, the $s=1,...,m$ primal and dual scaling functions can be defined as
\begin{align}
    \varphi^{\Gamma^*}_{s, m} (\mathbf{p}^*) &= \chi_{\mathbf{P}^*_s} (\mathbf{p}^*);  \nonumber\\
    \Tilde{\varphi}^{\Gamma^*}_{s, m} (\mathbf{p}^*) &= \delta (\mathbf{p}^*- \mathbf{p}^*_{v^*_s}),  \nonumber     
\end{align}
where $\chi_{\mathbf{P}^*_s}$ is the characteristic function defined on the $s$-th block of the partitioning, and $\delta (\mathbf{p}^*- \mathbf{p}^*_{v^*_s})$ is the Dirac delta with the energy centred on the point $\mathbf{p}^*_{v^*_s}$. (The $\Gamma^*$ superscript indicates the scaling functions are defined on the metrized line graph space.) The advantage of the characteristic functions is that they can be used to reveal the geometric information of the partitionings.
We also consider an alternative setup, namely
\begin{align}
    \label{equ:LG-primal-scaling-Kronecker}
    \varphi^{\Gamma^*}_{s, m} (\mathbf{p}^*) &= \delta_{\mathbf{p}^*_{v^*_s}, \mathbf{p}^*};\\
    \label{equ:LG-dual-scaling-delta}
    \Tilde{\varphi}^{\Gamma^*}_{s, m} (\mathbf{p}^*) &= \delta (\mathbf{p}^*- \mathbf{p}^*_{v^*_s}),   
\end{align}
where $\delta_{\mathbf{p}^*_{v^*_s}, \mathbf{p}^*}$ is the Kronecker delta. The construction using equations  (\ref{equ:LG-primal-scaling-Kronecker}) and (\ref{equ:LG-dual-scaling-delta}) can be considered akin to the lazy wavelets introduced by \cite{sweldens1996lifting}, but on the new vertex set.

Then the initial expansion form for the function approximation can be written as
\begin{align}\nonumber
\label{equ:LG-function-approx-stage-m}
    g^{\mathcal{E}}(e) \equiv g^{\mathcal{V}^*}(v^*) = g^{\Gamma^*}(\mathbf{p}^*)  
    = \sum_{s=1}^{m} c^{\Gamma^*}_{s, m} \varphi^{\Gamma^*}_{s, m} (\mathbf{p}^*)
\end{align}
where $\mathbf{p}^* \in \Gamma^*$ and $g^{\Gamma^*}$ is the metrized analogue of the function $g^{\mathcal{V}^*}$, and the (initial) scaling coefficients are $c^{\Gamma^*}_{s, m} := \langle g^{\Gamma^*}, \Tilde{\varphi}_{s,m}^{\Gamma^*} \rangle = g_s^{\Gamma^*} = g_s^{\mathcal{E}}$. 
From this point onwards, we use $g^{\mathcal{V}^*}$ interchangeably with $g^{\Gamma^*}$.
The aim is to find the function approximation as follows
\begin{equation}
\label{equ:final_expansion}
    g^{\Gamma^*}(\mathbf{p}^*) = \sum_{s \in \mathcal{S}_2} c^{\Gamma^*}_{s, 2} \varphi^{\Gamma^*}_{s, 2}(\mathbf{p}^*) + \sum_{l \in \mathcal{D}_2} d^{\Gamma^*}_{l} \psi^{\Gamma^*}_{l}(\mathbf{p}^*),
\end{equation}
where $\mathcal{D}_2 = \{k_m,...,k_3\}$, and $\mathcal{S}_2 = \{1,...,m\}\backslash \mathcal{D}_2$.
The set $\{d^{\Gamma^*}_{l}\}_{l \in \mathcal{D}_2}$ holds the detail (wavelet) coefficients, which can be obtained at each stage-$r$ along with the scaling coefficients $\{c^{\Gamma^*}_{s,r}\}_{s\in \mathcal{S}_r}$ by our proposed algorithm next detailed.

\subsection{LG-LOCAAT algorithm}
We now present our proposed algorithm in terms of the iterative split-predict-update-relink procedure on the set of graph edge observations 
$\{g^{\mathcal{E}}_k\}_{k=1}^m$. As discussed above, the edge observations can be mapped as 
$\{g^{\Gamma^*}_k\}_{k=1}^m$ into the metrized line graph domain.
As with LOCAAT \citep{jansen2009multiscale}, we start from stage-$m$, corresponding to the original edge-based observations, and the initial scaling coefficients are defined as $c^{\Gamma^*}_{k,m} := \langle g^{\Gamma^*}, \Tilde{\varphi}^{\Gamma^*}_{k,m} \rangle = g^{\Gamma^*}_k = g^{\mathcal{E}}_k$. The initial neighbourhood structure can be represented in the line graph domain as described in equations (\ref{equ:LG-neighbour1}) or (\ref{equ:LG-neighbour2}). In what follows, we write $\mathcal{N}^{\mathcal{V}^*}_{k,m}$ instead of $\mathcal{N}^{\mathcal{V}^*}_k$ since the neighbourhood structure will be changed as the algorithm progresses through stage-$m$, stage-$(m-1)$, and so on. 

\begin{itemize}
    \item \textbf{Split:} In line with \cite{jansen2009multiscale}, choose the new vertex to be removed, denote it by $v^*_{k_m}$, according to minimum integral value for the primal scaling function, here associated with its metrized point $\mathbf{p}^*_{v^*_{k_m}}$. When the initial primal scaling functions are defined as characteristic functions on the partitionings of the metrized line graph, the initial integral values are
    \begin{align}
        I^{\Gamma^*, \text{sum}}_{k,m} &= \int_{\Gamma^*} \varphi^{\Gamma^*}_{k, m}(\mathbf{p}^*) d\mathbf{p}^*   \nonumber \\
        &= \int_{\Gamma^*} \chi_{\mathbf{P}^*_k} (\mathbf{p}^*) d\mathbf{p}^*  \nonumber \\
        &= \mu(\mathbf{P}^*_k)   \nonumber \\
        \label{equ:sum-of-distance}
        &\propto \sum_{s: v^*_s \in \mathcal{N}^{\mathcal{V}^*}_{k,m}} \text{dist}_{\mathcal{V}^*}(v^*_k, v^*_s),
    \end{align}
    where we use the Lebesgue measure $\mu$ for a union of intervals (or line segments) to be the summation of their lengths, and `$\propto$' means `proportional to'. The resulting integral is proportional to the sum of distances of the chosen new vertex to its neighbouring vertices. Computationally, using a proportional sum of distances will result in the same detail coefficients as when using the sum of distances, as shown by the following proposition, whose proof appears in Appendix \ref{proof:proposition-LOCAAT-integral}.
    \begin{proposition}
    \label{pro:LOCAAT-integral}
        Suppose we have an integral sequence $\underline{I}^* = \{I^*_{k,m}\}_{k=1}^m$, and a constant $C>0$. Then, performing the LOCAAT algorithm with $C \cdot \underline{I}^*$ as integrals will yield the same detail coefficients and the same prediction/update filters, as performing LOCAAT with $\underline{I}^*$.
    \end{proposition}
    
    An alternative integral initialisation for the  scaling functions was suggested by \cite{jansen2009multiscale} to be the average distance, obtained as
    \begin{equation}
        \label{equ:average-distance}
        I^{\Gamma^*, \text{ave}}_{k,m} = \frac{1}{2 |\mathcal{N}^{\mathcal{V}^*}_{k,m}|} \sum_{s: v^*_s \in \mathcal{N}^{\mathcal{V}^*}_{k,m}} \text{dist}_{\mathcal{V}^*}(v^*_k, v^*_s). 
    \end{equation}
    We will also test the performance of the algorithm using the average distances as initial integrals since previous literature indicates its good performance \citep{mahadevan2010multiscale}.
    
    When the initial primal scaling functions are set as Kronecker delta, this leads to a variant of the lazy wavelets introduced in \cite{sweldens1998lifting}. 
    Denoting $\underline{\delta}_s = (0,\cdots, 0, 1, 0, \cdots, 0)$ as a canonical basis representation for $v^*_s$, such that only its $s$-th element is non-zero, we have 
    \begin{align}
        I^{\Gamma^*, \text{Delta}}_{k,m} = \langle \underline{\delta}_s, \mathbbm{1}_m \rangle  
        \label{equ:Kronecker-delta-integral}
        = 1,
    \end{align}
    where $\mathbbm{1}_m$ is a vector of ones of length $m$.
    
    Thus we will consider for the split step the following three integral value choices: sum of distances (equation (\ref{equ:sum-of-distance}), `sum'), average distance (equation (\ref{equ:average-distance}), `ave'), and starting with a vector of ones (equation (\ref{equ:Kronecker-delta-integral}), `Delta'). From now on, we will skip the superscript indicating the integral determination (sum/ave/Delta) unless necessary.
    
    Once the initial integral values have been decided, we choose the new vertex to be predicted and then removed corresponding to the minimum integral value. If there exist multiple new vertices with the minimum integral value, then we randomly pick one of these vertices. 
    As the LOCAAT framework follows the principle of recursive construction, we only need to fix the initial integrals and the recursive integral computation will be carried out through the iterative process, see \cite{sweldens1998lifting}. Therefore, in what follows, we refer to a general stage-$r$ and its associated removal index, $k_r$. 
    

    \item \textbf{Predict:} 
    The detail coefficient obtained at stage-$r$ is
    \begin{align}
        \label{equ:LG-predict}
        d^{\Gamma^*}_{k_r} = c^{\Gamma^*}_{k_r,r}- \sum_{s: v^*_s \in \mathcal{N}^{\mathcal{V}^*}_{k_r, r}} a^{\Gamma^*}_{s,r} c^{\Gamma^*}_{s, r},
    \end{align}
    where $\{a^{\Gamma^*}_{s,r}\}_{s: v^*_s \in \mathcal{N}^{\mathcal{V}^*}_{k_r, r}}$ are the prediction weights.
    The MRA framework associated with the prediction step can be recursively expressed as \citep{jansen2009multiscale}, 
    \begin{align}
        \Tilde{\psi}^{\Gamma^*}_{k_r} = \Tilde{\varphi}^{\Gamma^*}_{k_r} - \sum_{s: v^*_s \in \mathcal{N}^{\mathcal{V}^*}_{k_r, r}} a^{\Gamma^*}_{s,r} \Tilde{\varphi}^{\Gamma^*}_{s, r}.  \nonumber
    \end{align}
    Integrating and letting the left hand side be zero, the prediction weights satisfy
    \begin{align}
        \sum_{s: v^*_s \in \mathcal{N}^{\mathcal{V}^*}_{k_r, r}} a^{\Gamma^*}_{s,r} = 1.  \nonumber
    \end{align}
    Then, adapting the suggestion of \cite{jansen2009multiscale}, we construct the prediction weights by means of the normalised inverse distances,  
    \begin{align}
    \label{equ:inverse-distance-prediction}
        a^{\Gamma^*}_{s,r} = \frac{1/\text{dist}(\mathbf{p}^*_{v^*_{k_r}}, \mathbf{p}^*_{v^*_{s}})}{\sum_{t: v^*_t \in \mathcal{N}^{\mathcal{V}^*}_{k_r, r}} 1/\text{dist}(\mathbf{p}^*_{v^*_{k_r}}, \mathbf{p}^*_{v^*_{t}})}.
    \end{align}
    In addition to the inverse distance weights, we also consider a simple moving average prediction, where the prediction weight is 
    \begin{align}
    \label{equ:moving-average}
        a^{\Gamma^*}_{s,r} = \frac{1}{|\mathcal{N}^{\mathcal{V}^*}_{k_r, r}|}.
    \end{align}

    \item \textbf{Update:} The update step is performed for the integrals and scaling coefficients associated with the neighbourhood $\{s: v^*_{s} \in \mathcal{N}^{\mathcal{V}^*}_{k_r, r}\}$, 
    \begin{align}
        I^{\Gamma^*}_{s, r-1} &= I^{\Gamma^*}_{s, r} + a^{\Gamma^*}_{s,r} I^{\Gamma^*}_{k_r, r};  \nonumber\\
    \label{equ:LG-update}
        c^{\Gamma^*}_{s, r-1} &= c^{\Gamma^*}_{s,r} + b^{\Gamma^*}_{s,r} d^{\Gamma^*}_{k_r}, 
    \end{align}
    with the update filter $\{b^{\Gamma*}_{s,r}\}$ obtained by the minimum norm solution  as suggested in \cite{jansen2009multiscale}. Namely,
    \begin{equation}
    \label{equ:LG-LOCAAT-update}
        \text{$b^{\Gamma^*}_{s,r} = \frac{I^{\Gamma^*}_{s,r-1} I^{\Gamma^*}_{k_r, r}}{\sum_{t: v^*_t \in \mathcal{N}^{\mathcal{V}^*}_{k_r, r}} (I^{\Gamma^*}_{t,r-1})^2}$, \, for $s: v^*_{s} \in \mathcal{N}^{\mathcal{V}^*}_{k_r, r}$}.
    \end{equation}
    
    These update coefficients along with the condition $\sum_{s: v^*_s \in \mathcal{N}^{\mathcal{V}^*}_{k_r, r}} a^{\Gamma^*}_{s,r} = 1$, guarantee the stability of the transform, see \cite{jansen2005second} and \cite{jansen2009multiscale} for more details.

    \item \textbf{Relink:} A further necessity through the lifting steps is to relink the graph structure, since the removal of a vertex (and of its associated edges) might disconnect the graph structure. 
    For a tree graph, it is easy to see that the tree will be disconnected by removing a vertex which is not on the boundary (the boundary of a graph consists of those vertices with only one neighbouring vertex). However, for a non-tree graph (graphs satisfying $|\mathcal{E}| \geq |\mathcal{V}|$), the remaining subgraph after removing a vertex and associated edges might still be connected. For example, Figure \ref{fig:mlabel1} is a toy network from \cite{knight2019generalised}.
    Its graph structure is not that of a tree, and its structure will result in two separate components if we remove the 1-st vertex and its associated edges, but the remaining graph will still be connected if we remove the 2-nd vertex and its associated edges.
    
    \begin{figure}
    \centering
    \includegraphics[scale=0.33]{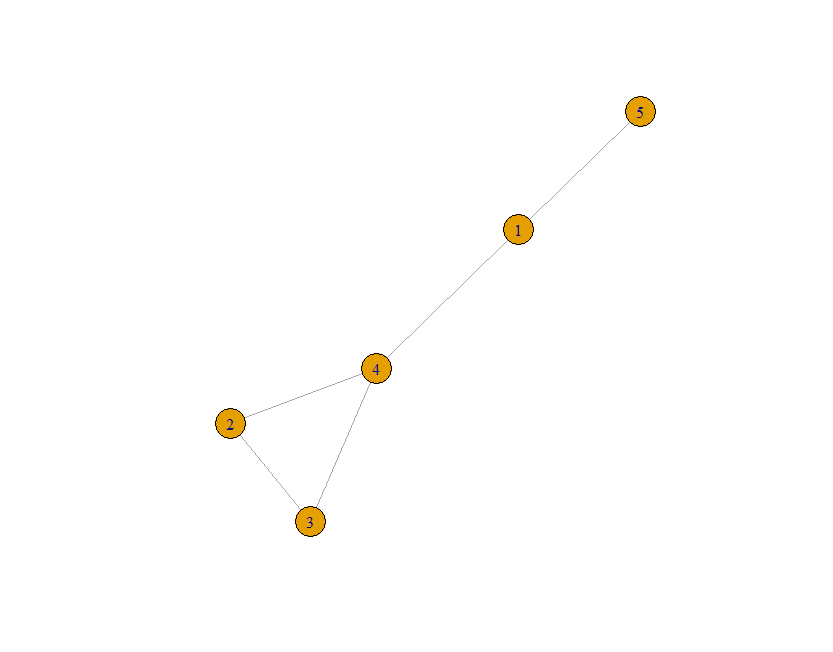}
    \caption{An undirected network structure `fiveNet' with five nodes from \cite{knight2019generalised}.}
    \label{fig:mlabel1}
    \end{figure} 

    Suppose we are at stage-$r$, then the relinkage part of the algorithm is carried out as follows.
    \begin{enumerate}
        \item Remove $v^*_{k_r}$ and all edges $\{e^*_l \,| \, v^*_{k_r} \in e^*_l \}$.
        \item Test the connectivity of the subgraph consisting of $\mathcal{N}^{\mathcal{V}^*}_{k_r, r}$ as vertices. If connected, then the relinkage will not be performed.
        \item If the subgraphs are not connected, then find the minimum spanning tree and embed this spanning tree into the existing graph structure.
    \end{enumerate}
    Once we complete the relinkage, the neighbourhood structure will be updated to be $\mathcal{N}^{\mathcal{V}^*}_{k, r-1}$ for all $k \in \{1,...,m\}\backslash \{k_m,...,k_r\}$. We denote the new graph structure at stage-$r$ as $G^*_{r} = (\mathcal{V}^*_r, \mathcal{E}^*_{r})$, and the one after relinkage as $G^*_{r-1} = (\mathcal{V}^*_{r-1}, \mathcal{E}^*_{r-1})$, where $\mathcal{V}^*_r = \mathcal{V}^*_{r-1} \cup \{v^*_{k_r}\}$.

    \item \textbf{Iterate:} We iterate the split-predict-update steps discussed above to obtain a sequence of detail coefficients $\{d^{\Gamma^*}_{k_m},..., d^{\Gamma^*}_{k_{\tau+1}}\}$, where $\tau$ is the stopping time indicating the number of new vertices that will not be removed. In our work, we set $\tau = 2$ as recommended in the literature \citep{jansen2009multiscale, nunes2006adaptive}. For details on the LOCAAT stopping time problem, the reader can refer to \cite{mahadevan2010multiscale}.
    \end{itemize}

    \noindent{\bf {Inverse LG-LOCAAT.}} A lifting scheme such as the one discussed above is a linear transform and guarantees a perfect reconstruction \citep{sweldens1998lifting}. Thus, the inverse transform is carried out by \textit{undoing} the lifting steps in equations (\ref{equ:LG-predict}) and (\ref{equ:LG-update}), from stage-$(r-1)$ to stage-$r$, 
    \begin{align}
        \label{equ:undo_update}
        c^{\Gamma^*}_{s, r} &= c^{\Gamma^*}_{s,r-1} - b^{\Gamma^*}_{s,r} d^{\Gamma^*}_{k_r}, \,\,\,\text{for $s: v^*_{s} \in \mathcal{N}^{\mathcal{V}^*}_{k_r, r}$},  \\
        \label{equ:undo_predict}
        c^{{\Gamma^*}}_{k_r, r} &= d^{\Gamma^*}_{k_r} + \sum_{s: v^*_s \in \mathcal{N}^{\mathcal{V}^*}_{k_r, r}} a^{\Gamma^*}_{s,r} c^{\Gamma^*}_{s, r}.
    \end{align}

    \noindent{\bf {Function expansion.}} Through the iterations of the LG-LOCAAT steps, after stage-$r$ we have
\begin{equation}
\label{equ:stage_r_expansion}
    g^{\Gamma^*}(\mathbf{p}^*) = \sum_{s \in \mathcal{S}_{r-1}} c^{\Gamma^*}_{s, r-1} \varphi^{\Gamma^*}_{s, r-1}(\mathbf{p}^*) + \sum_{l \in \mathcal{D}_{r-1}} d^{\Gamma^*}_{l} \psi^{\Gamma^*}_{l}(\mathbf{p}^*),
\end{equation}
where $\mathcal{D}_{r-1} = \{k_m,...,k_r\}$ are the wavelet coefficient indices, and $\mathcal{S}_{r-1} = \{1,...,m\}\backslash \mathcal{D}_{r-1}$ are the scaling indices. 
    
    \noindent{\bf {Computational viewpoint.}} A lifting coefficient array has to be stored after every stage to allow for the inverse transform to be carried out. 
    For a tree structure at stage-$(r-1)$, we start with $G^*_{r-1}$ and the set of edges $\{e^*_k = \{v^*_i, v^*_j\} \, | \, \text{$e^*_k \in \mathcal{E}^*_{r-1}$ and $v^*_i , v^*_j \in \mathcal{N}^{\mathcal{V}^*}_{k_r, r}$}\}$, then add the vertex $v^*_{k_r}$ and connect it to all vertices in $\mathcal{N}^{\mathcal{V}^*}_{k_r, r}$, thus obtaining $G^*_r$. For non-tree cases, it is possible that we do not disconnect some of the edges in the graph $G^*_{r-1}$ as discussed in relinkage. Therefore, we have to preserve more information than for the lifting array in \cite{jansen2009multiscale}, with components
    \begin{equation}
        k_r \quad \mathcal{N}^{\mathcal{V}^*}_{k_r, r} \quad S^*_{r} \quad \underline{a}^{\Gamma^*}_r \quad \underline{b}^{\Gamma^*}_r,  \nonumber
    \end{equation}
    where $S^*_{r}$ is the set consists of all $s$ such that $v^*_s \in \mathcal{N}^{\mathcal{V}^*}_{k_r, r}$, and $\underline{a}^*_r$, $\underline{b}^*_r$ are the prediction and update filters.
    In addition, 
    for stage-$r$, all pairs $(s, s')$, where $s\neq s'$ and $s,s' \in \mathcal{N}^{\mathcal{V}^*}_{k_r, r}$, such that $\{v^*_{s}, v^*_{s'}\} \in \mathcal{E}^*_{r}$ are added. 
    
Theoretical considerations for the LG-LOCAAT construction appear in Appendix~\ref{app:theory}, covering its sparsity, stability and scale definition.

\section{Simulation study}\label{sec:simstudy}
In this section we investigate the behaviour of our proposed LG-LOCAAT in the context of a nonparametric regression problem carried out over the edge set of a network. We model the edge observations as
\begin{align}
\label{equ:nonparametric-line-graph}
    f^\mathcal{E}_{k} &= g^\mathcal{E}(e_k) + \epsilon_k,  \nonumber \\
    &= g^\mathcal{E}_k + \epsilon_k. 
\end{align}
where $g^\mathcal{E}$ is the true, unknown function defined on the edge set $\mathcal{E} = \{e_k\}_{k=1}^m$, and $\{\epsilon_k\}_{k=1}^m$ is iid noise, assumed to follow a normal distribution $N(0, \sigma^2)$. Thus, $\underline{f}^\mathcal{E} = \{f^\mathcal{E}_k\}_{k=1}^m$ is the set of observation values on the edges of our graph $G$ that are corrupted by noise, and our aim is to obtain an estimator $\hat{g}^{\mathcal{E}}$ of the true (unknown) function $g^{\mathcal{E}}$ evaluated at the observed edges. 

The simulation study will assess three aspects indicative of the LG-LOCAAT's performance: stability (via the condition number), compression ability (via sparsity plots), and denoising performance (via the average mean squared error and associated estimator bias and variance). 

\subsection{Denoising strategy}\label{subsec:denoising}
Recall that the line graph transform allows us to represent $g^\mathcal{E}(e_k) = g^{\mathcal{V}^*}(v^*_k)$ (and similarly for the observations $\{f^\mathcal{E}_k\}_{k=1}^m$). Thus, equation (\ref{equ:nonparametric-line-graph}) can be rewritten as
\begin{align}
    f^{\mathcal{V}^*}_{k} &= g^{\mathcal{V}^*}(v^*_k) + \epsilon_k  \nonumber \\
    &= g^{\mathcal{V}^*}_k + \epsilon_k,  \nonumber
\end{align}
and we perform the LG-LOCAAT decomposition for the observations $\underline{f}^{\mathcal{V}^*}$, in order to obtain a sequence of detail coefficients $\underline{d}^*$. Next, the detail coefficients are separated into artificial levels by taking a quantile split of the integral values, see \cite{jansen2009multiscale} and \cite{nunes2006adaptive}. 
Since in practice the noise has an unknown variance which has to be first estimated before applying thresholding, we follow \cite{donoho1994ideal} and estimate $\sigma$ by the median absolute deviation (MAD) of the detail coefficients belonging to the finest artificial level. 

Wavelet thresholding will be performed on the detail coefficients. Following the finding from \cite{mahadevan2010multiscale}, which is that performing thresholding for around 80-90\% detail coefficients seems an optimal choice, we perform the wavelet thresholding for all detail coefficients except for those that have been allocated at the two coarsest artificial levels. We employ empirical Bayes thresholding,  where the non-zero part of the prior density is modelled as the `quasi-Cauchy' distribution from \cite{johnstone2004needles}, already proven to have good performance for LOCAAT-based approaches, see \cite{nunes2006adaptive} and \cite{jansen2009multiscale}.

A set of estimated coefficients, $\hat{\underline{{d}}}^*$, will be obtained following thresholding and we then perform the inverse LG-LOCAAT transform in order to obtain an estimate of the true, unknown function. We denote the corresponding estimates as $\hat{g}^{\mathcal{V}^*}_k$ for all $k\in \{1,...,m\}$ and we recall their equivalence to those in the original graph domain, such that $\hat{g}^{\mathcal{E}}_k = \hat{g}^{\mathcal{V}^*}_k$.

We investigate the effect of three different noise magnitudes, as measured by the signal-to-noise ratio (SNR = 3, 5, and 7). This ratio is given by $\text{SNR} = \sqrt{\text{var}(g^{\mathcal{E}})}/ \sigma$, where $\text{var}(g^{\mathcal{E}})$ is the variance of the simulated true function, and $\sigma$ is the standard deviation of the noise. For a clear comparison, the true function will be normalised so that $\text{var}(g^\mathcal{E}) = 1$. 

We sample $q=1,...,Q$ different graph structures, $G^{(q)}$. 
For each of them, we simulate $r=1,...,R$ different noise sequences, corresponding to a certain noise level, as discussed above. Following the estimation procedure, we calculate the average mean squared error (AMSE) defined as
\begin{align}
    \text{AMSE} = (Q R m)^{-1} \sum_{q=1}^{Q} \sum_{r=1}^R \sum_{k=1}^m \left(\hat{g}_{k,q,r}^{\mathcal{E}} - g_{k,q}^{\mathcal{E}} \right)^2,  \nonumber
\end{align}
where $g_{k,q}^{\mathcal{E}}$ is the true edge function value corresponding to the $k$-th `new' vertex on the line graph $G^{*(q)}$, while $\hat{g}_{k,q,r}^{\mathcal{E}}$ is the estimate of $g_{k,q}^{\mathcal{E}}$ when the true function is corrupted by the $r$-th noise sequence. 

We also investigate the variance and squared bias associated with our methods, where
\begin{align}
    \text{Var}
    = ({QRm})^{-1} \sum_{q=1}^Q\sum_{k=1}^m\sum_{r=1}^R \left( \hat{g}^{\mathcal{E}}_{k, q,r} - \Bar{g}^{\mathcal{E}}_{k,q} \right)^2,  \nonumber
\end{align}
where $\Bar{g}_{k,q} = \frac{1}{R}\sum_{r=1}^R \hat{g}^{\mathcal{E}}_{k,q,r}$ and the squared bias is calculated as 
\begin{align}
    \text{Bias}^2 = ({Qm})^{-1} \sum_{q=1}^Q \sum_{k=1}^m \left( \Bar{g}^{\mathcal{E}}_{k,q} -  g^{\mathcal{E}}_{k,q} \right)^2.  \nonumber
\end{align}

The true function $g^\mathcal{E}$ is generated by the test functions as described next. 

\subsection{Test Functions}
\label{Sec:test_funcs}
\begin{figure}
    \centering
    \includegraphics[width=6cm, height=10cm]{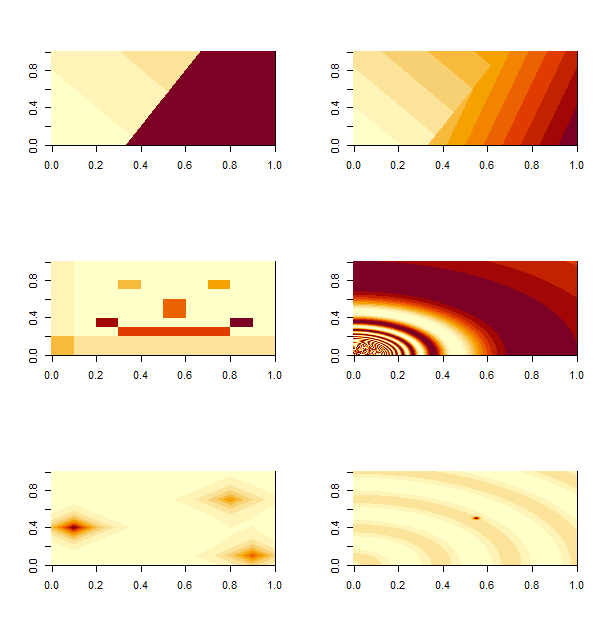}
    \caption{Test functions heat maps. From left to right on {\em top row}: $g_1$, maartenfunc (mfc); {\em middle row}: Blocks, Doppler; {\em bottom row}: Bumps, Heavisine.}
    \label{fig:test_funcs}
\end{figure}

The set of functions from \cite{jansen2009multiscale} along with the $g_1$ function introduced in \cite{mahadevan2010multiscale} will be used in our simulation study in order to capture the variety of traits that may occur in signals recorded over the edges of networks pertaining to fields from hydrology and transportation to computer networks. 
Figure \ref{fig:test_funcs} illustrates that `Blocks' is piecewise continuous with several discontinuities; `Doppler' is a smooth function; `Bumps' displays several spikes; `Heavisine' is a smooth function but with high variability; `{\tt mfc}' presents two discontinuous sections, while the $g_1$ has a similar but finer structure.

\subsubsection{Sampling Network Structure}
Since the test functions are defined over the square $[0,1]\times [0,1]$ (the formulae can be found in \cite{mahadevan2010multiscale}), we sample this space by means of a network structure. We first sample $n$ points $\{(x_i, y_i)\}_{i=1}^n$, where $x_i, y_i \sim \text{Unif}(0,1)$. These points are fixed as the graph vertices $\{v_i = (x_i, y_i)\}_{i=1}^n$ of the network $G$. The graph edges are obtained by the minimum spanning tree, which gives us a set of $m$ $(m = n-1)$ connections between vertices. We let these be the set of edges for the graph $G$, with the length of each edge given by the Euclidean distance between the two vertices associated with this edge. 

\begin{figure}
    \centering
    \includegraphics[width=8cm, height=8cm]{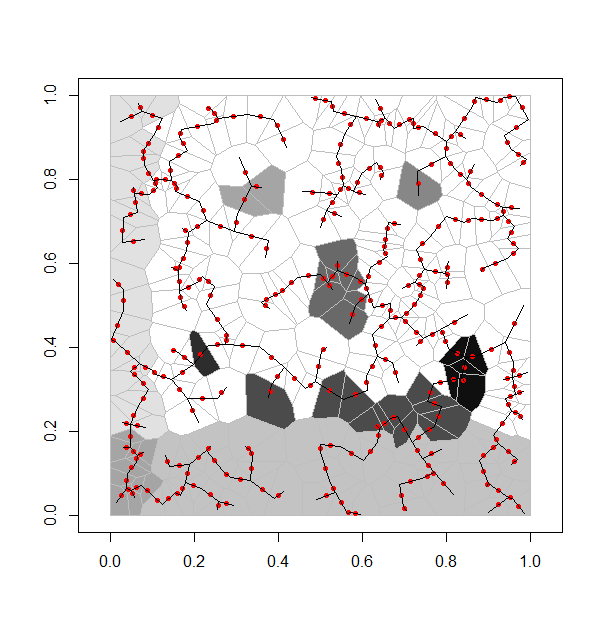}
    \caption{Pointwise Blocks function. Each Voronoi polygon represents the function value at the corresponding new vertex (original edge), represented by the red points.}
    \label{fig:visual_normal_funcs}
\end{figure}

An example appears in Figure \ref{fig:visual_normal_funcs}, displaying the network vertices and edges superimposed with the Blocks function, evaluated as next described and visualised using a Voronoi polygon tessellation centred at the edge midpoints.

\begin{figure}
    \centering
    \includegraphics[width=8cm, height=8cm]{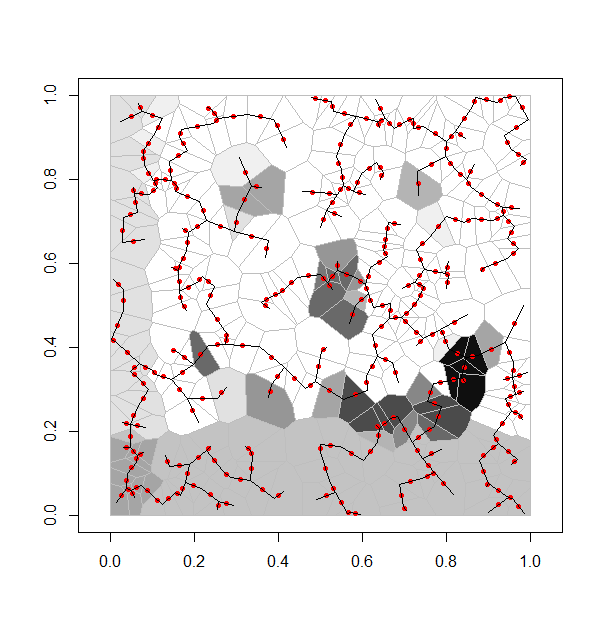}
    \caption{Edge averaging Blocks function. Each Voronoi polygon represents the function value at the corresponding new vertex (original edge), represented by the red points.}
    \label{fig:visual_edge_average}
\end{figure}

\subsubsection{Embedding the Function Values}
Within our study, we evaluate the function values over the network edges using two methods. The first method is to simply select the value that corresponds to the coordinate at the midpoint of each edge. Let us assume we have a function $g^{\text{test}}$ from the previously mentioned collection of test functions. Consequently, for an edge $e_{k} = \{v_i, v_j\}$, where $v_i = (x_i, y_i)$ and $v_j = (x_j, y_j)$, the corresponding `true' observation value will be
\begin{align}
\label{equ:test_func_normal}
    g^{\mathcal{E}}_k  = g^{\mathcal{V}^*}_k := g^{\text{test}}(\frac{x_i + x_j}{2}, \frac{y_i + y_j}{2}). 
\end{align}
Nonetheless, this definition still hinges on a vertex-wise perspective and we refer to these as `pointwise' functions. We additionally explore an alternative method for designing the function values, which incorporates the network geometry. Inspired by the cell averaging of \cite{donoho1997cart}, we view the edges as line segments and for $e_{k} = \{v_i, v_j\}$, we define the function value as
\begin{align}
\label{equ:test_func_edge_average}
    g^{\mathcal{E}}_k &= g^{\mathcal{V}^*}_k:= \nonumber\\ 
    &= \frac{1}{N}\sum_{h=0}^{N-1} g^{\text{test}}\left(x_i +  \frac{x_j - x_i}{N-1} h , y_i +  \frac{y_j - y_i}{N-1} h\right). 
\end{align}
We refer to these as `edge averaging' functions. 
Here we let $N=100$. Figures \ref{fig:visual_normal_funcs} and \ref{fig:visual_edge_average} show the `pointwise' and `edge averaging' Blocks functions, respectively. Note that when compared with the pointwise function, the edge averaging is smoother when the edges cross blocks with different values.

\subsection{Results}\label{subsec:simresults}
In this section, we will provide numerical results for the proposed LG-LOCAAT algorithm which gives rise to several variants determined by specific combinations of initial scaling function integral definition and prediction weights, as well as the availability of coordinate information or only of the path length. 
The acronyms we use to identify the approaches we explore, as well as their brief descriptions, appear in Table \ref{tab:LG_acronym}.

\begin{table*}[hbt!]
    \centering
    \begin{tabular}{|c| p{12cm}|}
    \hline
    Acronym & Proposed LG-LOCAAT variant  \\
    \hline
    LG-Sid-c & S: sum of distances as integral (equation (\ref{equ:sum-of-distance})); id: inverse distance prediction (equation (\ref{equ:inverse-distance-prediction})); c: coordinate information available.  \\
    \hline
    LG-Aid-c & A: average distance as integral (equation (\ref{equ:average-distance})); id: inverse distance prediction (equation (\ref{equ:inverse-distance-prediction})); c: coordinate information available.  \\
    \hline
    LG-Did-c & D: a sequence of ones as integrals (equation (\ref{equ:Kronecker-delta-integral})); id: inverse distance prediction (equation (\ref{equ:inverse-distance-prediction})); c: coordinate information available.  \\
    \hline
    LG-Snw-c & S: sum of distances as integral (equation (\ref{equ:sum-of-distance})); nw: moving average prediction (equation (\ref{equ:moving-average})); c: coordinate information available.  \\
    \hline
    LG-Anw-c & A: average distance as integral (equation (\ref{equ:average-distance})); nw: moving average prediction (equation (\ref{equ:moving-average})); c: coordinate information available. \\
    \hline
    LG-Dnw-c & D: a sequence of ones as integrals (equation (\ref{equ:Kronecker-delta-integral})); nw: moving average prediction (equation (\ref{equ:moving-average})); c: coordinate information available. \\
    \Xhline{2pt}
    LG-Sid-p & S: sum of distances as integral (equation (\ref{equ:sum-of-distance})); id: inverse distance prediction (equation (\ref{equ:inverse-distance-prediction})); p: path length available.  \\
    \hline
    LG-Aid-p & A: average distance as integral (equation (\ref{equ:average-distance})); id: inverse distance prediction (equation (\ref{equ:inverse-distance-prediction})); p: path length available.  \\
    \hline
    LG-Did-p & D: a sequence of ones as integrals (equation (\ref{equ:Kronecker-delta-integral})); id: inverse distance prediction (equation (\ref{equ:inverse-distance-prediction})); p: path length available.  \\
    \hline
    LG-Snw-p & S: sum of distances as integral (equation (\ref{equ:sum-of-distance})); nw: moving average prediction (equation (\ref{equ:moving-average})); p: path length available.  \\
    \hline
    LG-Anw-p & A: average distance as integral (equation (\ref{equ:average-distance})); nw: moving average prediction (equation (\ref{equ:moving-average})); p: path length available.  \\
    \hline
    LG-Dnw-p & D: a sequence of ones as integrals (equation (\ref{equ:Kronecker-delta-integral})); nw: moving average prediction (equation (\ref{equ:moving-average})); p: path length available.  \\
    \hline 
    \end{tabular}
    \vspace{0.3cm}
    \caption{Acronyms and algorithm descriptions for different parameter choices of LG-LOCAAT.}
    \label{tab:LG_acronym}
\end{table*}

\subsubsection{Stability}
\label{sec:simulation_LG_condition_number}
As the lifting scheme is a linear transform, it can be represented using matrix multiplication, $\underline{d} = \Tilde{R} \underline{g}$, where $\underline{g}$ is the observation sequence, $\Tilde{R}$ denotes the forward matrix generated by lifting scheme (see \cite{nunes2006adaptive} for details on its LOCAAT construction), and $\underline{d}$ is the detail coefficient sequence. The condition number of the lifting matrix will be used to assess the stability of the transform and can be computed as 
$
    \kappa (\Tilde{R}) = {\rho_1}/{\rho_m},
$
where $\{\rho_i\}_{i=1}^m$ is the set of corresponding singular values ordered according to their magnitude, such that $\rho_1$ is the maximum singular value and $\rho_m$ is the minimum one. The condition number satisfies $\kappa (\Tilde{R}) \geq 1$ and the lower its value, the more stable the transform \citep{higham2002accuracy} is. 

Tables \ref{tab:LG_condition_number} and \ref{tab:LG_condition_number_without}  illustrate that all algorithms are characterised by similar condition numbers, with a preference for initial integrals as a sequence of ones and moving average prediction weights.

\subsubsection{Sparsity}

The tool we use to assess the performance of compression is the sparsity plot, constructed as follows. First, we perform the LG-LOCAAT transform on the sequence of true values $\{g^{\mathcal{V}^*}_{k}\}_{k=1}^{m}$. This will yield a corresponding vector which contains two scaling coefficients and $(m-2)$ detail coefficients. We begin with just two scaling coefficients and assume all detail coefficients are zero, then perform the inverse transform for this vector. In general, obtain an estimator $\hat{g}^{\mathcal{V}^*}(t)$ for the true function $g^{\mathcal{V}^*}$, where $(t-1)$ gives the number of detail coefficients used in reconstruction, such that $\hat{g}^{{\mathcal{V}^*}}(1)$ is the reconstruction with only two scaling coefficients, while $\hat{g}^{{\mathcal{V}^*}}(2)$ means the reconstruction with two scaling coefficients and one detail coefficient with the largest absolute value, and so on. For any step, we will introduce the detail coefficient with the largest absolute value into the vector for reconstruction. We thus obtain a different estimate at each step, until all detail coefficients have been used. For the sparsity plot, the $x$-axis is the `$t$'-argument and the value on the $y$-axis is the integrated squared error (ISE), defined as 
\begin{equation*}
    \text{ISE}(t) = Q^{-1} \sum_{q=1}^{Q} \sum_{k=1}^{m} \left(\hat{g}^{{\mathcal{V}^*}^{(q)}}_{k}(t) - g^{{\mathcal{V}^*}^{(q)}}_{k}\right)^2,
\end{equation*}
where $g^{{\mathcal{V}^*}^{(q)}}_k$ and $\hat{g}^{{\mathcal{V}^*}^{(q)}}_k$ denote the true observation at new vertex $v^*_k$ and its reconstruction uses $(t-1)$ detail coefficients for the $q$-th network. 
If the ISE decays to zero fast (if for small $t$, the ISE is already close to zero), then the algorithm leads to a sparse decomposition for the target function.


\noindent{\bf {Results for Pointwise Functions.}} Figure \ref{fig:sparsity_coord_normal_func} shows the sparsity results across the (pointwise) test functions, for schemes based on the coordinate information. Note that the compression ability of the algorithm on $g_1$ and {\tt mfc} surpasses that on any other test function. Bumps and Doppler functions display similar sparsity results, while for Blocks the results are inferior. The compression of Heavisine is the weakest compared with other functions. Reassuringly, there is no significant difference among the choices of integral values and prediction weights. 
Figure \ref{fig:sparsity_path_normal_func} shows the results for the same functions while using path length instead of the coordinates. For data compression, there is no significant difference between using path length and coordinate information.


\noindent{\bf {Results for Edge Averaging Functions.}} Figure \ref{fig:sparsity_coord_edge_average} shows the sparsity results for different edge averaging functions. The compression ability follows the similar patterns to pointwise functions. Along with Figure \ref{fig:sparsity_path_edge_average}, again there is no evidence of a significant difference between using path length and coordinate information. Both sets of plots indicate that different choices of integral values and prediction weights will not significantly affect the algorithm's compression ability.

\subsubsection{Denoising Performance}
In this section, we investigate the behaviour of LG-LOCAAT in the context of the network edge nonparametric regression problem outlined in Section \ref{subsec:denoising} using $R=100$ noise sequences over $Q=50$ graph structures each with $m=99$ edges. 

\noindent{\bf{Results for Pointwise Functions.}}
In terms of AMSE, Table \ref{tab:edge_LG_AMSE_99-coordinate} illustrates that using the average distance as the integral value surpasses the choice of sequence of ones or sum integral choices. Methods `Did' and `Dnw' perform well for most of the function except Blocks and Heavisine, while `Sid' and `Snw' never appear to be the optimal choice unless the function is the Blocks (piecewise constant). From Table \ref{tab:edge_LG_variance_99-coordinate}, we note that the integral choice `A' (average distance) provides the best results for variance control, while the choice `S' (sum of distances) provides similar and competitive results, too. Performing the algorithm with the integral choice `D' (sequence of ones) gives a relatively high variance compared with the other two choices. The reason for this might be the dual wavelet functions have fewer overlaps than any other scaling function construction choices (recall that the integral choice `D' allows us to capture the information more uniformly on the graph structure). For the bias results in Table \ref{tab:edge_LG_bias_99-coordinate}, the integral choice `D' surpasses the other methods for most of the functions. The choice `A' is competitive for the Heavisine function, which contains high frequency components when compared with other functions. The choice `S' introduces high bias for the smoother functions (Doppler, Heavisine, {\tt mfc}). Hence, when using coordinate information, overall `LG-Aid-c' appears to be a balanced choice that delivers competitive results irrespective of signal smoothness and noise contamination level. Although `LG-Did-c' does not give as good results as `LG-Aid-c' in terms of AMSE (especially for Blocks and Heavisine) and variance, it is still worth emphasising on since it yields a low bias. Let us next investigate the impact of not accessing the coordinate information.

Performing the algorithm by using the path distance instead of coordinate information, the AMSE results are remarkably comparable to those when coordinate information is available, see Table \ref{tab:edge_LG_AMSE_99-path}. The variance and bias patterns in Tables \ref{tab:edge_LG_variance_99-path} and \ref{tab:edge_LG_bias_99-path} appear similar to previous methods. However, should coordinate information be available, this may be the better choice. This may be justified by the design of the function values based on a two-dimensional Euclidean space and it will be essential to inspect the algorithm performance for the edge averaging functions (see next section). 

Additionally, the average distance for integral and the inverse distance prediction (`LG-Aid-p') also arise as a strong choice in this context, with the `Did' choice a close competitor in terms of bias control, and `Sid' a close match particularly for variance results. The performance of `LG-Sid-p' is very similar to `LG-Aid-p' except for Heavisine. So using `average distances' as integral is advantageous when dealing with high-frequency information, as opposed to using `sum of distances' as integral. However, we note that the choice of average distances as integral values lacks theoretical support.

\begin{table*}[hbt!]
    \centering
    \scalebox{1}{
    \begin{tabular}{|c|c|c|c|c|c|c|c|c|}
    \hline
    AMSE$\times 10^3$ (sd$\times 10^3$) & $g_1$ & Blocks & Doppler & Bumps & Heavisine & mfc \\
    \hline
    \multicolumn{6}{c}{SNR=3}    \\
    \hline
    LG-Sid-c & 65 (19) & 114 (44) & 109 (39) & 92 (26) & 272 (99) & 52 (12) \\
    \hline
    \textbf{LG-Aid-c} & \textbf{62} (18) & \textbf{113} (48) & \textbf{92} (29) & \textbf{79} (19) & \textbf{198} (66) & 46 (10) \\
    \hline
    LG-Did-c & 64 (18) & 120 (46) & 93 (31) & 80 (20) & 235 (81) & \textbf{45} (11) \\
    \hline
    LG-Snw-c & 66 (19) & 117 (47) & 109 (38) & 97 (28) & 282 (103) & 53 (12) \\
    \hline
    LG-Anw-c & 64 (18) & 116 (48) & \textbf{92} (31) & 81 (20) & 205 (66) & 46 (10) \\
    \hline
    LG-Dnw-c & 67 (19) & 126 (51) & 98 (33) & 83 (21) & 261 (87) & 47 (11) \\
    \hline
    \multicolumn{6}{c}{SNR=5}    \\
    \hline
    LG-Sid-c & 23 (7) & \textbf{41} (16) & 44 (17) & 45 (17) & 206 (88) & 26 (6) \\
    \hline
    \textbf{LG-Aid-c} & 23 (7) & 42 (18) & 38 (13) & \textbf{38} (12) & \textbf{138} (48) & \textbf{22} (5) \\
    \hline
    LG-Did-c & 23 (7) & 46 (20) & \textbf{37} (13) & 39 (12) & 178 (76) & \textbf{22} (5) \\
    \hline
    LG-Snw-c & \textbf{22} (7) & 42 (17) & 44 (17) & 47 (18) & 217 (91) & 26 (6) \\
    \hline
    LG-Anw-c & 23 (8) & 43 (19) & 38 (14) & 39 (12) & 147 (48) & \textbf{22} (5) \\
    \hline
    LG-Dnw-c & 24 (8) & 48 (21) & 39 (14) & 41 (13) & 208 (81) & 23 (5) \\
    \hline
    \multicolumn{6}{c}{SNR=7}    \\
    \hline
    LG-Sid-c & 11 (3) & \textbf{21} (9) & 24 (10) & 28 (13) & 184 (83) & 17 (4) \\
    \hline
    \textbf{LG-Aid-c} & 11 (3) & 22 (10) & \textbf{21} (7) & \textbf{23} (8) & \textbf{120} (43) & \textbf{14} (3) \\
    \hline
    LG-Did-c & 11 (3) & 24 (11) & \textbf{21} (7) & 24 (9) & 161 (75) & \textbf{14} (3) \\
    \hline
    LG-Snw-c & \textbf{10} (3) & \textbf{21} (9) & 25 (10) & 29 (14) & 195 (86) & 18 (4) \\
    \hline
    LG-Anw-c & \textbf{10} (3) & 22 (12) & 22 (8) & 24 (8) & 130 (43) & \textbf{14} (3) \\
    \hline
    LG-Dnw-c & 11 (3) & 25 (11) & 22 (8) & 25 (9) & 191 (79) & 15 (3) \\
    \hline
    \end{tabular}}
    \vspace{0.3cm}
    \caption{AMSE for LG-LOCAAT on a tree structure with 100 nodes and 99 edges. The functions follow the pointwise construction. We assume the coordinate information is available. The values in parentheses are the standard deviations ($\times 10^3$) of the AMSE results across the $Q \times R = 50 \times 100$ replications.}
    \label{tab:edge_LG_AMSE_99-coordinate}
\end{table*}

\begin{table*}[hbt!]
    \centering
    \scalebox{1}{
    \begin{tabular}{|c|c|c|c|c|c|c|c|c|}
    \hline
    AMSE$\times 10^3$ (sd$\times 10^3$) & $g_1$ & Blocks & Doppler & Bumps & Heavisine & mfc \\
    \hline
    \multicolumn{6}{c}{SNR=3}    \\
    \hline
    LG-Sid-p & 67 (21) & \textbf{116} (45) & 109 (39) & 93 (25) & 274 (96) & 54 (12) \\
    \hline
    \textbf{LG-Aid-p} & \textbf{65} (20) & 117 (48) & \textbf{94} (31) & 84 (21) & \textbf{209} (73) & \textbf{47} (10) \\
    \hline
    LG-Did-p & \textbf{65} (19) & 122 (49) & 95 (31) & \textbf{83} (22) & 253 (93) & \textbf{47} (11) \\
    \hline
    LG-Snw-p & 68 (21) & 118 (48) & 110 (36) & 96 (26) & 288 (102) & 54 (12) \\
    \hline
    LG-Anw-p & 66 (20) & 124 (55) & 95 (31) & 84 (22) & 215 (76) & 48 (10) \\
    \hline
    LG-Dnw-p & 68 (20) & 129 (54) & 100 (33) & 86 (22) & 276 (93) & 48 (11) \\
    \hline
    \multicolumn{6}{c}{SNR=5}    \\
    \hline
    LG-Sid-p & \textbf{23} (8) & \textbf{42} (17) & 44 (17) & 45 (16) & 208 (83) & 27 (7) \\
    \hline
    \textbf{LG-Aid-p} & \textbf{23} (8) & 44 (19) & \textbf{38} (13) & \textbf{40} (13) & \textbf{152} (62) & \textbf{23} (5) \\
    \hline
    LG-Did-p & 24 (8) & 47 (21) & \textbf{38} (13) & \textbf{40} (13) & 192 (86) & \textbf{23} (5) \\
    \hline
    LG-Snw-p & \textbf{23} (8) & \textbf{42} (19) & 44 (16) & 47 (17) & 220 (87) & 27 (6) \\
    \hline
    LG-Anw-p & 24 (8) & 46 (22) & 39 (13) & \textbf{40} (13) & 159 (61) & \textbf{23} (5) \\
    \hline
    LG-Dnw-p & 25 (8) & 49 (22) & 40 (14) & 42 (13) & 219 (89) & 24 (5) \\
    \hline
    \multicolumn{6}{c}{SNR=7}    \\
    \hline
    LG-Sid-p & \textbf{11} (3) & 22 (10) & 24 (10) & 28 (12) & 186 (78) & 18 (5) \\
    \hline
    \textbf{LG-Aid-p} & \textbf{11} (3) & 23 (11) & \textbf{21} (7) & \textbf{24} (9) & \textbf{135} (58) & \textbf{15} (3) \\
    \hline
    LG-Did-p & \textbf{11} (4) & 25 (12) & \textbf{21} (7) & 25 (9) & 172 (83) & \textbf{15} (3) \\
    \hline
    LG-Snw-p & \textbf{11} (3) & \textbf{21} (10) & 25 (10) & 29 (13) & 197 (80) & 18 (5) \\
    \hline
    LG-Anw-p & \textbf{11} (3) & 24 (13) & 22 (8) & \textbf{24} (9) & 142 (55) & \textbf{15} (3) \\
    \hline
    LG-Dnw-p & \textbf{11} (4) & 26 (12) & 22 (8) & 26 (10) & 201 (87) & 16 (4) \\
    \hline
    \end{tabular}}
    \vspace{0.3cm}
    \caption{AMSE for LG-LOCAAT on a tree structure with 100 nodes and 99 edges. The functions follow the pointwise construction. The path distance is used. The values in parentheses are the standard deviations ($\times 10^3$) of the AMSE results across the $Q \times R = 50 \times 100$ replications.}
    \label{tab:edge_LG_AMSE_99-path}
\end{table*}

\noindent{\bf{Results for Edge Averaging Functions.}} Tables \ref{tab:edge_average_LG_AMSE_99-coordinate} and \ref{tab:edge_average_LG_AMSE_99-path} illustrate that when employing our algorithm on edge averaging functions, the `Aid' algorithm is the optimal choice in terms of the AMSE when the coordinate information is known, while `Did' performs better if only path lengths are known. However, `Aid' surpasses the other methods for high frequency Heavisine function. Choosing a sequence of ones as the starting integrals decreases the bias for all functions, and the improvements are more significant when we use the path length instead of the coordinate information (Tables \ref{tab:edge_average_LG_bias_99-coordinate} and \ref{tab:edge_average_LG_bias_99-path}). Again, `Aid' is the optimal choice in terms of variance control  (Tables \ref{tab:edge_average_LG_variance_99-coordinate} and \ref{tab:edge_average_LG_variance_99-path}). 
The methods with coordinate information still slightly surpass the corresponding results by the path distance, except for mfc. 
According to AMSE, the performances of `Did' and `Aid' are close except when the underlying function has high frequency (e.g., Heavisine). 

Overall, we recommend the use of coordinate information when available, as implemented in the methods, LG-Aid-c and LG-Did-c, if the underlying function are spatial coordinate dependent. Should such coordinate information not be available, LG-Aid-p and LG-Did-p are also very competitive throughout the board.

\begin{table*}[hbt!]
    \centering
    \scalebox{1}{
    \begin{tabular}{|c|c|c|c|c|c|c|c|c|}
    \hline
    AMSE$\times 10^3$ (sd$\times 10^3$) & $g_1$ & Blocks & Doppler & Bumps & Heavisine & mfc \\
    \hline
    \multicolumn{6}{c}{SNR=3}    \\
    \hline
    LG-Sid-c & 64 (19) & 126 (47) & 103 (32) & 92 (25) & 281 (96) & 51 (12) \\
    \hline
    \textbf{LG-Aid-c} & \textbf{59} (16) & \textbf{120} (48) & 89 (27) & \textbf{79} (18) & \textbf{206} (70) & 45 (10) \\
    \hline
    \textbf{LG-Did-c} & \textbf{59} (16) & \textbf{120} (43) & \textbf{88} (26) & \textbf{79} (19) & 228 (77) & \textbf{44} (10) \\
    \hline
    LG-Snw-c & 65 (19) & 128 (46) & 104 (32) & 96 (27) & 292 (101) & 52 (12) \\
    \hline
    LG-Anw-c & 60 (16) & 125 (46) & 90 (28) & 80 (19) & 215 (70) & 45 (10) \\
    \hline
    LG-Dnw-c & 61 (16) & 128 (46) & 92 (28) & 82 (20) & 255 (83) & 46 (11) \\
    \hline
    \multicolumn{6}{c}{SNR=5}    \\
    \hline
    LG-Sid-c & 26 (9) & 50 (22) & 45 (17) & 45 (16) & 220 (86) & 25 (6) \\
    \hline
    \textbf{LG-Aid-c} & \textbf{24} (8) & \textbf{49} (23) & 39 (14) & \textbf{38} (11) & \textbf{147} (53) & \textbf{21} (4) \\
    \hline
    LG-Did-c & 25 (7) & 51 (21) & \textbf{38} (14) & \textbf{38} (12) & 176 (71) & \textbf{21} (4) \\
    \hline
    LG-Snw-c & 26 (9) & 51 (22) & 46 (18) & 47 (17) & 230 (90) & 26 (6) \\
    \hline
    LG-Anw-c & 25 (9) & 51 (24) & 40 (14) & 39 (12) & 158 (53) & 22 (5) \\
    \hline
    LG-Dnw-c & 26 (8) & 54 (23) & 40 (14) & 40 (12) & 204 (76) & 22 (5) \\
    \hline
    \multicolumn{6}{c}{SNR=7}    \\
    \hline
    LG-Sid-c & \textbf{13} (4) & \textbf{27} (12) & 26 (11) & 28 (12) & 199 (81) & 17 (4) \\
    \hline
    \textbf{LG-Aid-c} & \textbf{13} (4) & \textbf{27} (13) & \textbf{22} (8) & \textbf{23} (8) & \textbf{129} (48) & 14 (3) \\
    \hline
    LG-Did-c & \textbf{13} (4) & 28 (13) & \textbf{22} (8) & 24 (8) & 160 (69) & \textbf{13} (3) \\
    \hline
    LG-Snw-c & \textbf{13} (4) & \textbf{27} (12) & 26 (11) & 29 (13) & 211 (86) & 18 (5) \\
    \hline
    LG-Anw-c & \textbf{13} (5) & \textbf{27} (15) & 23 (9) & 24 (8) & 141 (47) & 14 (3) \\
    \hline
    LG-Dnw-c & \textbf{13} (5) & 30 (14) & 23 (8) & 25 (9) & 190 (74) & 15 (3) \\
    \hline
    \end{tabular}}
    \vspace{0.3cm}
    \caption{AMSE for LG-LOCAAT on a tree structure with 100 nodes and 99 edges. The functions follow the edge-averaging construction. We assume the coordinate information is available. The values in parentheses are the standard deviations ($\times 10^3$) of the AMSE results across the $Q \times R = 50 \times 100$ replications.}
    \label{tab:edge_average_LG_AMSE_99-coordinate}
\end{table*}

\begin{table*}[hbt!]
    \centering
    \scalebox{1}{
    \begin{tabular}{|c|c|c|c|c|c|c|c|c|}
    \hline
    AMSE$\times 10^3$ (sd$\times 10^3$) & $g_1$ & Blocks & Doppler & Bumps & Heavisine & mfc \\
    \hline
    \multicolumn{6}{c}{SNR=3}    \\
    \hline
    LG-Sid-p & 66 (20) & 127 (47) & 104 (32) & 93 (25) & 282 (95) & 53 (12) \\
    \hline
    LG-Aid-p & 62 (17) & 123 (47) & 91 (28) & 83 (21) & \textbf{218 (76)} & \textbf{46 (10)} \\
    \hline
    \textbf{LG-Did-p} & \textbf{60} (17) & \textbf{122} (45) & \textbf{90} (27) & \textbf{82} (21) & 247 (87) & \textbf{46} (11) \\
    \hline
    LG-Snw-p & 66 (21) & 129 (48) & 104 (30) & 96 (26) & 296 (101) & 53 (12) \\
    \hline
    LG-Anw-p & 62 (17) & 133 (51) & 92 (28) & 84 (22) & 225 (77) & 47 (10) \\
    \hline
    LG-Dnw-p & 62 (17) & 130 (49) & 94 (28) & 85 (22) & 270 (89) & 47 (11) \\
    \hline
    \multicolumn{6}{c}{SNR=5}    \\
    \hline
    LG-Sid-p & 26 (9) & \textbf{51} (23) & 45 (17) & 46 (16) & 222 (85) & 26 (7) \\
    \hline
    LG-Aid-p & 26 (9) & 52 (24) & \textbf{39} (14) & 40 (12) & \textbf{162 (64)} & \textbf{22 (5)} \\
    \hline
    \textbf{LG-Did-p} & \textbf{25} (8) & 52 (22) & \textbf{39} (14) & \textbf{39 (12)} & 190 (82) & \textbf{22 (5)} \\
    \hline
    LG-Snw-p & 26 (9) & 52 (23) & 46 (17) & 47 (17) & 233 (88) & 27 (6) \\
    \hline
    LG-Anw-p & 26 (9) & 55 (27) & 41 (14) & 40 (12) & 170 (63) & 23 (5) \\
    \hline
    LG-Dnw-p & 26 (8) & 55 (25) & 41 (15) & 41 (13) & 217 (85) & 23 (5) \\
    \hline
    \multicolumn{6}{c}{SNR=7}    \\
    \hline
    LG-Sid-p & \textbf{13} (5) & \textbf{27} (13) & 26 (11) & 28 (12) & 202 (81) & 18 (5) \\
    \hline
    \textbf{LG-Aid-p} & \textbf{13} (4) & 28 (15) & \textbf{22} (8) & \textbf{24} (9) & \textbf{145} (60) & \textbf{14} (3) \\
    \hline
    LG-Did-p & \textbf{13} (4) & 29 (14) & \textbf{22} (8) & \textbf{24} (9) & 173 (78) & \textbf{14} (3) \\
    \hline
    LG-Snw-p & \textbf{13}  (5) & \textbf{27} (13) & 26 (11) & 29 (12) & 212 (81) & 18 (5) \\
    \hline
    LG-Anw-p & 14 (5) & 29 (16) & 23 (8) & 25 (9) & 154 (57) & 15 (3) \\
    \hline
    LG-Dnw-p & \textbf{13} (5) & 31 (15) & 23 (9) & 26 (9) & 201 (83) & 16 (3) \\
    \hline
    \end{tabular}}
    \vspace{0.3cm}
    \caption{AMSE for LG-LOCAAT on a tree structure with 100 nodes and 99 edges. The functions follow the edge-averaging construction. The path distance is used. The values in parentheses are the standard deviations ($\times 10^3$) of the AMSE results across the $Q \times R = 50 \times 100$ replications.}
    \label{tab:edge_average_LG_AMSE_99-path}
\end{table*}


\section{Hydrological data analysis}\label{sec:realdata}
\begin{figure}
    \centering
    \includegraphics[width=8cm, height=8cm]{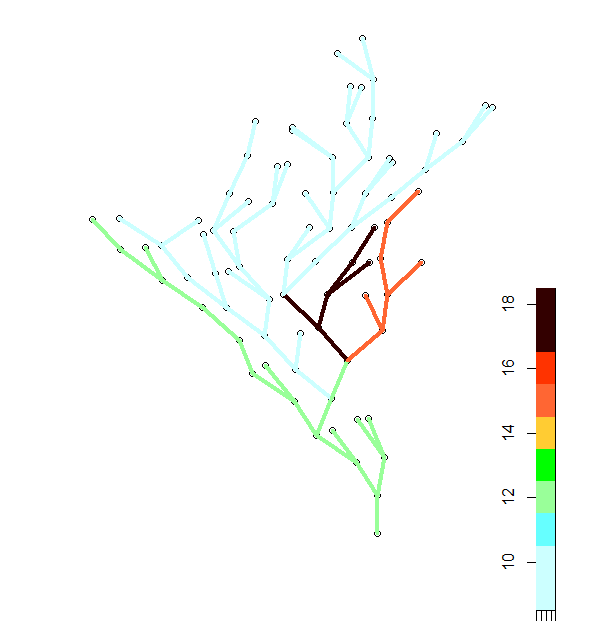}
    \caption{The simulated river flow data, the network structure is introduced in \cite{gallacher2017flow} and the test function construction is from \cite{park2022lifting}.}
    \label{fig:simulated_flow}
\end{figure}

In this section we tackle the dissolved oxygen (DO) dataset that motivated this work, where DO measures the amount of oxygen in water and acts as an indicator of water quality. In order to assess the behaviour of our proposed algorithms on hydrological data, we take an additional step through first analysing a piecewise constant function mimicking river flow, as introduced by \cite{park2022lifting} on a simulated tree network from \cite{gallacher2017flow}. This simulated flow data can be visualised in Figure \ref{fig:simulated_flow} and is also investigated as a real data-like example by \cite{park2022lifting}, who evaluate their proposed method on it.  

\subsection{Flow-based Function Denoising}
\label{sec:simulated-flow}
The river network contains 80 vertices and 79 edges, and its edge set is separated into seven different clusters. The flow function is constructed as follows \citep{park2022lifting}. The function values for every stream are initially set as 9, then a cluster is randomly picked and the function values of every stream in this cluster will be randomly chosen from the values $\{12,15,18\}$. We continue this procedure until there are more than 30 edges containing a value larger than 9. 
\begin{figure}
    \centering
    \includegraphics[width=8cm, height=8cm]{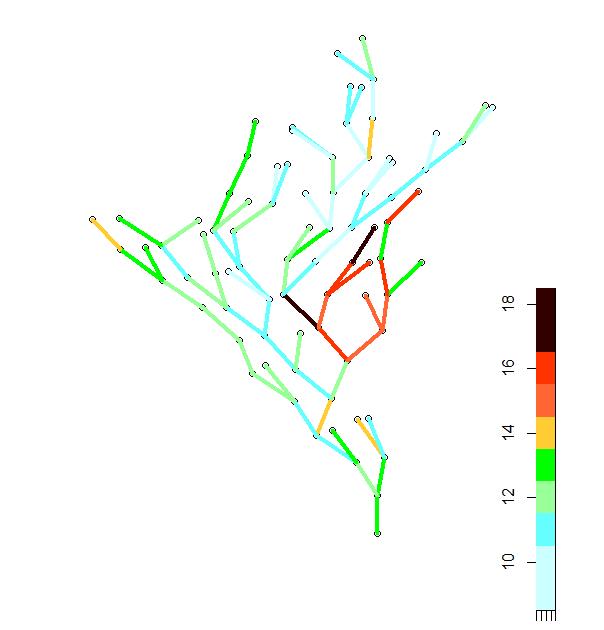}
    \caption{The flow data corrupted by noise $\epsilon \sim N(0,4)$.}
    \label{fig:noisy_flow}
\end{figure}
Figure \ref{fig:noisy_flow} shows the noise corrupted flow data, where the noise is $\epsilon\sim N(0, 4)$ \citep{park2022lifting}.

\subsubsection{Nondecimated lifting transform}\label{subsec:nlt}
For completeness, we formally bring the nondecimation concept into the graph data setup. \cite{knight2009nondecimated} introduced a nondecimated lifting transform (NLT) based on the LOCAAT framework from \cite{jansen2009multiscale} where instead of using the shift operator, their proposal was to use different `trajectories' of removal order. 

In the graph context here, where we have for an unknown function $g^{\mathcal{E}}: \mathcal{E} \longrightarrow \mathbb{R}$
the set of noisy observations $\{f^\mathcal{E}_{k}\}^m_{k = 1}$ on the graph edge set $\{e_{k}\}^m_{k=1}$, 
we propose to naturally follow the idea from \cite{knight2009nondecimated} and generate $P$ trajectories $\{T_p\}_{p=1}^P$, where $T_p = (e_{o_1}, ..., e_{o_m})$ and $(o_1,...,o_m)$ is a permutation of the ordered sequence $(1,...,m)$. For each trajectory, let us say $T_p$, a detail coefficient sequence $\underline{d}^{*, (p)}$ can be obtained by any of (as long as it is the same one) the proposed transforms. Then an estimator $\hat{g}^{\mathcal{E},(p)}$ will be obtained corresponding to each trajectory $T_p$, and we propose to use the average estimator 
\begin{align}
    \hat{\Bar{g}}^{\mathcal{E}}_k = \frac{1}{P} \sum_{p=1}^P \hat{g}^{\mathcal{E},(p)}_k,  \nonumber
\end{align}
for $k \in \{1,...,m\}$. In the work from \cite{knight2009nondecimated}, a genetic algorithm \citep{lucasius1993understanding,lucasius1994understanding} is applied to generate `well-behaved' trajectories, which are those likely to have low variations. In the context of applying the LOCAAT algorithm for river network applications, \cite{park2022lifting} suggest to perform several permutations within each stream cluster for each trajectory. In this work, we assume that there is no prior information on the underlying function, thus, the trajectories are chosen by random permutations, but note that further improvements may be possible as suggested by \cite{knight2009nondecimated} and \cite{park2022lifting}.

\subsubsection{Flow Denoising Results}
We now investigate our proposed algorithms and their nondecimated versions on the noise contaminated flow dataset. Specifically, we employ the `LG-Sid-p' and `LG-Aid-p' algorithms with the path length instead of using the coordinate information, since the proposals in \cite{park2022lifting} are also based on path length metric and therefore this ensures a fair comparison. 
Table \ref{tab:flow-based-AMSE} shows the results when using our algorithms (one trajectory, following the minimum integral paradigm) as well as their nondecimated versions (acronyms ending in `nlt' in Table \ref{tab:flow-based-AMSE}). For completeness, we also display the results of the methods from \cite{park2022lifting}.
\begin{table*}[hbt!]
    \centering
    \scalebox{0.8}{
    \begin{tabular}{|c|c|c|c|c|c|c|c|}
    \hline
    AMSE (Std. error) & 80 obs ($\sigma = 1$) & 80 obs ($\sigma = 1.5$) & 80 obs ($\sigma = 2$)  \\
    \hline
    LG-Sid-p  & 0.6637 (0.0937) & 0.9770 (0.1428) & 1.2651 (0.1844)  \\
    \hline
    LG-Sid-p-nlt (30)  & 0.5704 (0.0779) & 0.8132 (0.1127) & 1.0346 (0.1433)  \\
    \hline
    LG-Aid-p  & 0.7183 (0.1139) & 1.0484 (0.1480) & 1.3429 (0.1932)  \\
    \hline
    \textbf{LG-Aid-p-nlt (30)}  & \textbf{0.5645 (0.0786)} & \textbf{0.8050 (0.1144)} & \textbf{1.0231 (0.1456)}  \\
    \Xhline{2pt}
    Proposed (Median) from \cite{park2022lifting} & 0.7265 (0.1212) & 1.0249 (0.1510) &  1.2818 (0.2599)  \\
    \hline
    Proposed (Hard) from \cite{park2022lifting} & 0.7396 (0.1317) & 1.0666 (0.1678) & 1.3705 (0.2495)  \\
    \hline
    Proposed (Median, nlt) from \cite{park2022lifting} & 0.7162 (0.1219) & 1.0106 (0.1678) & 1.2816 (0.2712)  \\
    \hline
    O’Donnell & 1.2698 (0.1251) & 1.3815 (0.1727) & 1.5421 (0.2017)  \\
    \hline
    \end{tabular}
    }
    \vspace{0.3cm}
    \caption{AMSE (and associated standard error) for different methods performed on simulated flow data.}
    \label{tab:flow-based-AMSE}
\end{table*}




We observe that for the simulated flow data with the proposed trajectory choice, `LG-Sid-p' results surpass all other (decimated) results.
Notably, with only one choice of trajectory, we obtain better results than the nondecimated results (50 trajectories) from \cite{park2022lifting}, while NLT has been shown in the literature to drastically improve denoising results \citep{knight2009nondecimated}. 


The line graph algorithms work better than the other choices, especially `LG-Aid-p-nlt', which gives consistently most competitive results, regardless of the noise level, as (very closely) does `LG-Sid-p-nlt'. 

We also note here the usefulness of introducing the random trajectory choice via nondecimation through the use of averaged estimators. In particular, exploring 30 trajectories has a significant impact on lowering the AMSE especially for lower signal-to-noise ratios. For the highest reported noise level, the percentage improvement by introducing several trajectories is around 23.8\% for `LG-Aid-p-nlt'.
The percentage improvement decreases with the decreasing noise level, but it is still substantial for our methods, ranging from around 21.4\% for `LG-Aid-p-nlt' 
to the lowest improvement is for `LG-Sid-p-nlt' (although this notably has a more competitive starting value for one trajectory), nevertheless still sizeable at 14.1\%, 16.7\%, and 18.2\% corresponding to the reported noise level ordering in Table \ref{tab:flow-based-AMSE}.

\begin{figure}
    \centering
    \includegraphics[width=8cm, height=8cm]{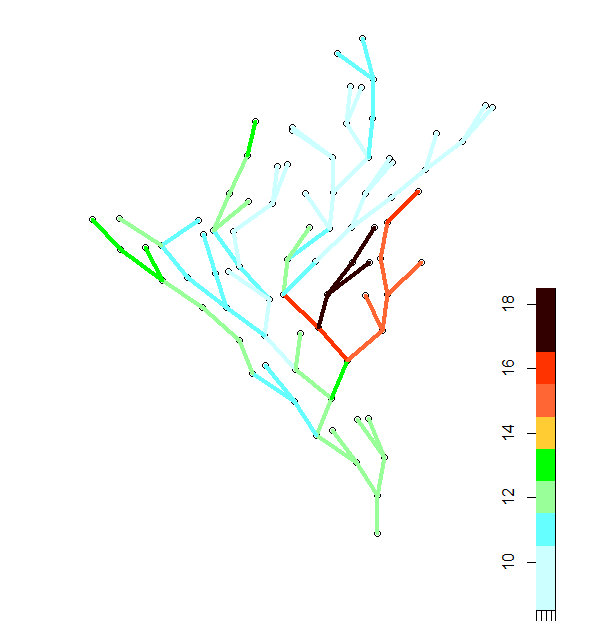}
    \caption{The denoised river flow data from Figure \ref{fig:noisy_flow} by means of `LG-Aid-p-nlt' with 30 trajectories.}
    \label{fig:denoised_flow}
\end{figure}

Figure \ref{fig:denoised_flow} shows the denoised signal by `LG-Aid-p-nlt', and we note that our algorithm is able to capture the true underlying 
structure when the noise level is high, demonstrating our algorithms work well for the hydrological data traits.

\subsection{Real Data Analysis}
In this section, we perform some of our proposed algorithms (`LG-Aid-nlt' and `LG-Did-nlt') on the DO dataset. 
 Recall we argued that it is natural to consider data collected from stations as edge-based observations, since each station is connected to a certain river body. In order to obtain the whole network structure, we also need to define the vertices. A natural choice would be to consider river conjunctions, the sources and mouths of rivers as the vertices, with each river being recognised as an edge. However, through this construction the England river network would be a tree structure with a large number of vertices, while only 60 observations are available. 
Even assuming the whole network structure has been constructed, some  information would still be difficult to obtain, for example, the length of an edge (river) since each river is a curve instead of a line segment. 

Our proposed line graph-based construction crucially enables us to bypass these problems and allows us the framework in which to model the stations as the line graph vertices.
In our context here, we propose to consider their associated latitude and longitude as their coordinates.

\noindent{\bf{Remark (Neighbourhood Structure).}}
In the work from \cite{park2022lifting}, the neighbourhood selection is based on the concept of `flow-connectedness' introduced by \cite{hoef2006spatial}. Under this concept, if the intersection of upstreams of two stations is a non-empty set, then they are defined as neighbours. In our work, we conjecture that river quality related indices are highly dependent on both river network structure and on localised human and animal behaviour. Hence, defining the neighbourhood structure as in equation (\ref{equ:LG-neighbour1}) and performing the LG-LOCAAT algorithm with coordinate information is useful since it enables us to capture these traits. 

\noindent{\bf{DO Dataset Results.}}
Guided by the results from previous simulation studies, we carry out our analysis by the denoising strategy described in Sections \ref{subsec:denoising} and \ref{subsec:nlt} using the nondecimated versions of our proposed `LG-Aid-c' and `LG-Did-c' for obtaining the detail coefficients. Since increasing the number of trajectories in nondecimated lifting increases the denoising performance, we perform a 100-trajectory nondecimated lifting on the DO dataset. 

\begin{figure}[H]
    \centering
    \includegraphics[width=6cm, height=7cm]{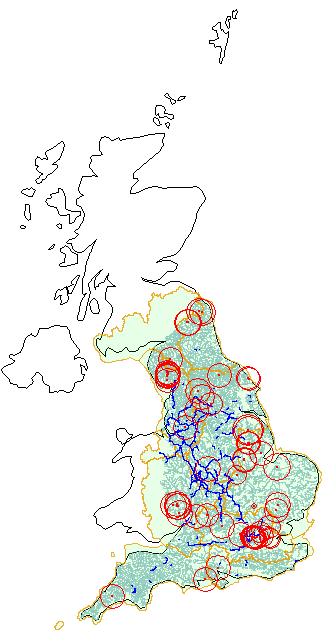}
    \caption{The denoised version of the DO data shown in Figure \ref{fig:UK_river_network} using `LG-Did-c-nlt' with 100 trajectories.}
    \label{fig:UK_river_denoised}
\end{figure}

\begin{figure}[H]
    \centering
    \includegraphics[width=6cm, height=7cm]{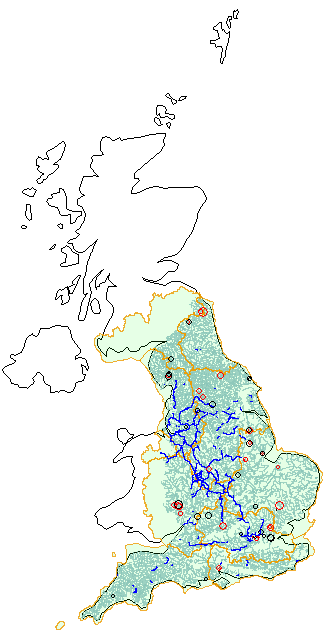}
    \caption{Residual visualisation. Positive residuals are represented by red circles, while negative residuals are represented by black circles. The circle size is determined by the residual absolute value.}
    \label{fig:UK_river_residual}
\end{figure}

Figure \ref{fig:UK_river_denoised} shows the denoised version of the DO data shown in Figure \ref{fig:UK_river_network}, while Figure \ref{fig:UK_river_residual} represents the residuals following our estimation. We observe that our methods tend to smooth the observations based on the network structure, see the red circle chain from  Wales to the southeast of England, and the values in northern England. Also note (Figure \ref{fig:UK_river_denoised}) that the DO values tend to be large around the countryside and areas close to the sea thus indicating good water quality, with low values around the London and Manchester areas, following the intuition that big cities have a negative impact on the water quality.

\begin{figure}
    \centering
    \includegraphics[width=8cm, height=6cm]{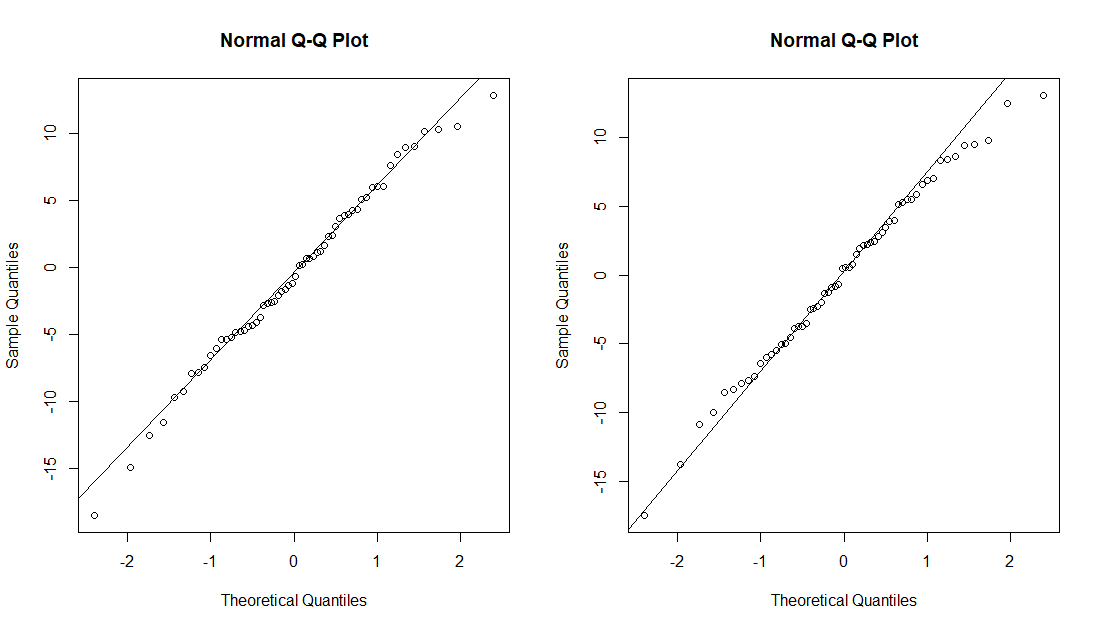}
    \caption{\textbf{Left:} Residual Q-Q plot of the DO data analysis (the data is collected on 10/May/2024, there are observations from all 60 stations) with proposed `LG-Aid-c-nlt'. \textbf{Right:} Residual Q-Q plot obtained using `LG-Did-c-nlt'. }
    \label{fig:10_May}
\end{figure}

\begin{figure}
    \centering
    \includegraphics[width=8cm, height=6cm]{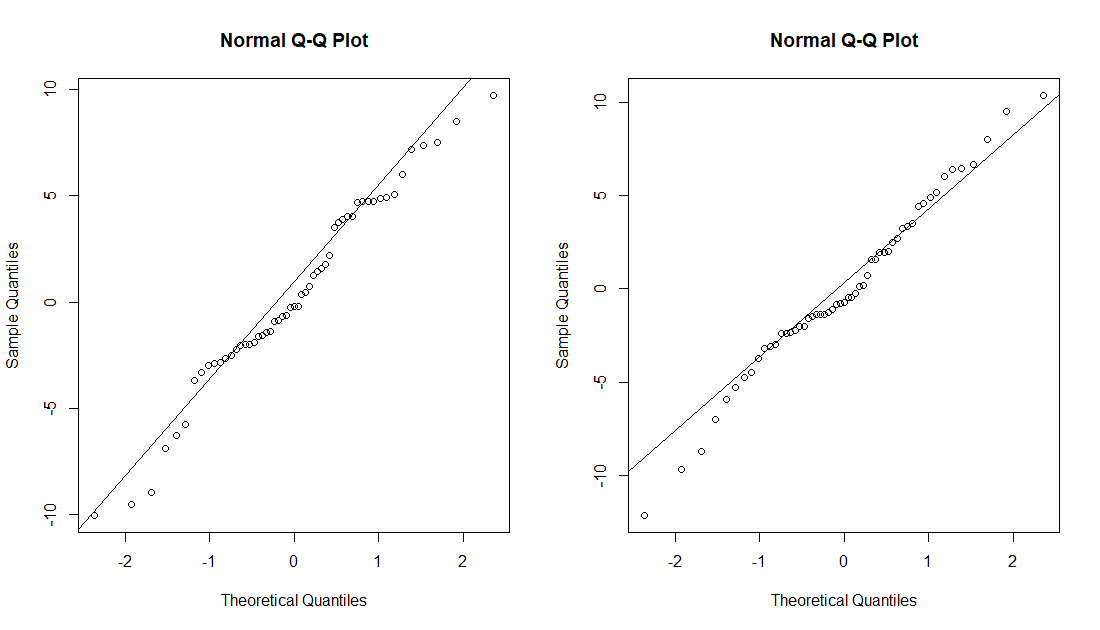}
    \caption{\textbf{Left:} Residual Q-Q plot of the DO data analysis (the data is collected on 05/Jun/2024, there are 55 observations from 60 stations with 5 missing observations) with proposed `LG-Aid-c-nlt'. \textbf{Right:} Residual Q-Q plot obtained using `LG-Did-c-nlt'. }
    \label{fig:05_Jun}
\end{figure}

Figures \ref{fig:10_May} and \ref{fig:05_Jun} show the residual Q-Q plots for the DO data analysis at two different time snapshots. The residuals for the data collected on 10/May/2024 appear to closely follow the normal distribution. The residual Q-Q plots for the data collected on 05/Jun/2024 are less well behaved especially for the `LG-Did' algorithm that uses a constant integral initialisation. This illustrates the underlying network impact on the quality of the estimate, since the existence of 5 missing values on that date resulted in having to connect stations that were not geographically close in order to ensure a connected network and additionally these features were not accounted for since the algorithm used a constant integral initialisation. 

On balance, these results give us the confidence that our algorithms work well for this dataset, and more generally the nondecimated `LG-Aid' appears to be the best choice for  hydrological data. 

\section{Conclusions and further work}\label{sec:concl}
Work on hydrological datasets \citep{park2022lifting} viewed as the analysis of information collected over a network (modelled as a graph), where the observations often pertain to the graph edges (e.g., rivers) rather than to its vertices (e.g., confluences), calls for specifically designed methodology in order to avoid the pitfalls of using noisy edge information as weights, for example. This paper has introduced a novel wavelet-like transformation, LG-LOCAAT,  designed to work on network edge-collected data, and has demonstrated its usefulness in the context of an application on denoising a water quality index. Future work could focus on addressing (i) limitations, such as the fact that we rely on a line graph transform with its associated increased complexity and potential inability to represent the data in its original domain at each stage of the algorithm, (ii) extensions such as, the development of a framework to allow for the computation of confidence intervals; new models able to incorporate time-dependency into the noise structure paralleling the GNAR-edge model of \cite{mantziou2023the}.



\bibliography{sn-bibliography}

\addtocounter{section}{0}
\appendixtitleon

\begin{appendices}

\section{Theoretical considerations}\label{app:theory}

\subsection{The rationale behind using Euclidean coordinates}
Many works in the literature explore the connection between graphs and manifolds \citep{bolker2002graph,singer2006graph}. 
\cite{zhou2008high} show that a graph can be considered as the discrete approximation of a manifold, or a manifold can be treated as the continuous analogue of a graph. 

\begin{figure}
    \centering
    \includegraphics[width=7.5cm]{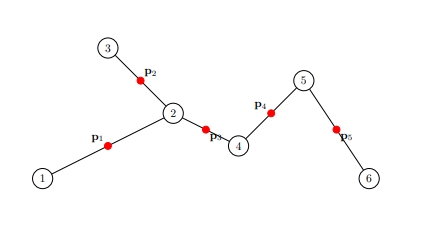}
    \caption{A toy tree graph with six vertices.}
    \label{fig:graph_manifold}
\end{figure}

A manifold is a topological space that is locally homeomorphic to an Euclidean space $\mathbb{R}^p$ for some $p$. Hence, any point on the manifold has a neighbourhood that can be mapped onto an open set of $\mathbb{R}^p$ \citep{tu2011manifolds}. Therefore, although a manifold is a space
naturally without coordinates, one may often use coordinate information for its local structures. Consider the toy graph in Figure \ref{fig:graph_manifold}, which contains six vertices $\{v_i\}_{i=1}^6$. The points $\{\mathbf{p}_k\}_{k=1}^5$ on the edges represent the metrized locations of the observations and we denote by $e_k$ the edge associated to $\mathbf{p}_k$ and by $U_k$ the open interval corresponding to the $k$-th edge (see Section \ref{sec:graph}). The open set associated to incident edges (for example, edges 1, 2 and 3) can be considered as a homeomorphism of a continuous open set in a Euclidean space. Hence, the Euclidean distance (for example, among $\mathbf{p}_1$, $\mathbf{p}_2$ and $\mathbf{p}_3$) supplies local structure information when performing the line graph transformation. 


\subsection{Sparsity}
In the context of wavelet transforms, the term `sparsity' usually indicates that the wavelet (detail) coefficients form a sparse sequence which represents the original signal energy by concentrating it in a small amount of coefficients. The properties of the signal are of course an essential factor in achieving sparsity, as well as the traits of the wavelet transform. In our context, recall our focus is on representing in the lifting domain a function $g^{\Gamma^*}$ defined on the metric space $(\Gamma^*, \text{dist})$, where the metrized locations $\{\mathbf{p}^*_{v^*_k}\}_{k=1}^m \in \Gamma^*$ are associated with the line graph vertices $\{v^*_{k}\}_{k=1}^m \in \mathcal{V}^*$.
Then a Euclidean analogue of Lipschitz continuity on a graph can be generalised as follows.
\begin{definition}
    A function $g^{\Gamma^*}$ defined on $\Gamma^*$ has a point $\mathbf{p}^*_{L}$ of Lipschitz continuity if there exists an interval $\mathbf{I}^* \subset\Gamma^*$, such that for all $\mathbf{p}^* \in \mathbf{I}^*$,
    \begin{align}
        \big|g^{\Gamma^*}(\mathbf{p}^*_{L}) - g^{\Gamma^*}(\mathbf{p}^*)\big| \leq C \, \text{dist}(\mathbf{p}^*_{L}, \mathbf{p}^*),  \nonumber
    \end{align}
    where $0< C <\infty$ is a constant.
\end{definition}
The spatial Lipschitz continuity used in \cite{jansen2009multiscale} and the H\"older class for tree graphs defined by \cite{gavish2010multiscale} motivate this definition, which allows a bound to be established on the detail coefficients, as described next.
\begin{proposition}
\label{pro:LG-decay}
    For a stage-$r$ prediction step, if for all $v^*_k \in \mathcal{N}^{\mathcal{V}^*}_{k_r,r} \cup \{v^*_{k_r}\}$, we have $c^{\Gamma^*}_{k, r} = g^{\Gamma^*}(\mathbf{p}^*_{v^*_{k}})$, and $g^{\Gamma^*}$ is Lipschitz continuous over an interval that contains  the metrized points associated with $\mathcal{N}^{\mathcal{V}^*}_{k_r,r}$, then the detail coefficient obtained by equation (\ref{equ:LG-predict}) at stage-$r$ satisfies that
    \begin{align}
    \label{equ:LG-decay}
        \left|d^{\Gamma^*}_{k_r}\right| &\leq C \, \frac{\sum_{s: v^*_s \in \mathcal{N}^{\mathcal{V}^*}_{k_r, r}} \text{dist}(\mathbf{p}^*_{v^*_{k_r}}, \mathbf{p}^*_{v^*_{s}})}{|\mathcal{N}^{\mathcal{V}^*}_{k_r, r}|}. 
    \end{align}
\end{proposition}
For the the proof, the reader can refer to Appendix \ref{app:proofs}.
Nonetheless, acquiring a precise bound for all detail coefficients is not a straightforward task for iterative methods.
Let us consider the following situation, suppose at stage-$r$, the function underpinning the scaling coefficients ($c^{\Gamma*}_{k, r}$) is Lipschitz continuous, which indicates that
\begin{align}
        \big|c^{\Gamma^*}_{k,r} - c^{\Gamma^*}_{s,r}\big| \leq C \, \text{dist}(\mathbf{p}^*_{v^*_{k_r}}, \mathbf{p}^*_{v^*_{s}}).  \nonumber
\end{align}
Assuming $v^*_s \in \mathcal{N}^{\mathcal{V}^*}_{k_r, r}$, and $v^*_k \notin \mathcal{N}^{\mathcal{V}^*}_{k_r, r}$, then from stage-$r$ to stage-$(r-1)$, we have 
\begin{align}
    c^{\Gamma^*}_{k,r-1} &= c^{\Gamma^*}_{k,r},  \nonumber \\
    c^{\Gamma^*}_{s,r-1} &= c^{\Gamma^*}_{s,r} + b^{\Gamma^*}_{s,r} d^{\Gamma^*}_{k_r}.  \nonumber
\end{align}
Under the assumptions of Proposition \ref{pro:LG-decay}, the bound for the absolute difference between $c^{\Gamma^*}_{k,r-1}$ and $c^{\Gamma^*}_{s,r-1}$ then becomes
\begin{align}
    \left|c^{\Gamma^*}_{k,r-1} - c^{\Gamma^*}_{s,r-1}\right| &= \left|\left(c^{\Gamma^*}_{k,r} - c^{\Gamma^*}_{s,r}\right) - b^{\Gamma^*}_{s,r} d^{\Gamma^*}_{k_r}\right|  \nonumber \\
    &\leq \big|c^{\Gamma^*}_{k,r} - c^{\Gamma^*}_{s,r}\big| + b^{\Gamma^*}_{s,r} \left|d^{\Gamma^*}_{k_r}\right|  \nonumber \\
    &\leq C \, \text{dist}(\mathbf{p}^*_{v^*_{k_r}}, \mathbf{p}^*_{v^*_{s}}) \nonumber \\ 
    &+ b^{\Gamma^*}_{s,r} C \, \frac{\sum_{t: v^*_t \in \mathcal{N}^{\mathcal{V}^*}_{k_r, r}} \text{dist}(\mathbf{p}^*_{v^*_{k_r}}, \mathbf{p}^*_{v^*_{t}})}{|\mathcal{N}^{\mathcal{V}^*}_{k_r, r}|}.  \nonumber
\end{align}
Hence, there is no guarantee that the function underpinning $c^{\Gamma^*}_{k, r-1}$ is still Lipschitz continuous with the same constant $C$. This indicates that even if $v^*_{k_r}$ and $v^*_{k_{r-1}}$ are `close' to each other for some stage-$r$, unwanted effects on the compression ability of the algorithm may still occur.

\subsection{Stability}
As wavelet functions generated by the lifting scheme (including LOCAAT-based approaches) are no longer orthogonal, the algorithm stability may become an issue. One way to guarantee the stability of the transform is to ensure that both primal and dual wavelet functions form Riesz bases, such that
\begin{align}
    L \, \lVert g^{\Gamma^*} \rVert_{L_2}^2 \leq &\sum_{k \in \mathcal{D}_r} |\langle g^{\Gamma^*}, \psi^{\Gamma^*}_{k} \rangle|^2  \leq U \, \lVert g^{\Gamma^*} \rVert_{L_2}^2,  \\
    \Tilde{L} \, \lVert g^{\Gamma^*} \rVert_{L_2}^2 \leq &\sum_{k \in \mathcal{D}_r} |\langle g^{\Gamma^*}, \Tilde{\psi}^{\Gamma^*}_{k} \rangle|^2  \leq \Tilde{U} \, \lVert g^{\Gamma^*} \rVert_{L_2}^2,
\end{align}
where $\mathcal{D}_r = \{k_m,...,k_{r+1}\}$.
If we can find the upper bounds, then the lower bound can be obtained by  duality, such that $L = {\Tilde{U}}^{-1}$ and $\Tilde{L} = U^{-1}$ holds, see \cite{cohen1992biorthogonal} and \cite{daubechies1992ten}.
However, as pointed out by \cite{jansen2009multiscale}, it is challenging to verify whether a set of bases satisfies the Riesz condition on a global scale, especially in irregular scenarios. \cite{simoens2003stabilized} and \cite{jansen2005second} presented a necessary but not sufficient condition for the Riesz condition is that each one-step transform and its inverse are uniformly bounded. This can be articulated as the one-level lifting operator and its inverse being bounded, see \cite{simoens2003stabilized}. For the LOCAAT-based algorithm, it means that the quantities resulting from the predict (equation (\ref{equ:LG-predict})), the update (equation (\ref{equ:LG-update})), undo update (equation (\ref{equ:undo_update})), and undo predict (equation (\ref{equ:undo_predict})) should be bounded in norm. For the forward prediction, given that $\sum_{s: v^*_k\in \mathcal{N}^{\mathcal{V}^*}_{k_r, r}} a^{\Gamma^*}_{s,r} = 1$ and $a^{\Gamma^*}_{s,r}\geq 0$, we have that 
\begin{align}
    |d^{\Gamma^*}_{k_r}|^2  &= |c^{\Gamma^*}_{k_r,r}- \sum_{s: v^*_s \in \mathcal{N}^{\mathcal{V}^*}_{k_r, r}} a^{\Gamma^*}_{s,r} c^{\Gamma^*}_{s, r}|^2  \nonumber \\
    \label{equ:predict_bound}
    &\leq (1 + \sum_{s: v^*_s\in \mathcal{N}^{\mathcal{V}^*}_{k_r, r}} |a^{\Gamma^*}_{s,r}|^2) \, \sum_{k: v^*_k\in \mathcal{N}^{\mathcal{V}^*}_{k_r, r} \cup \{k_r\}} |c^{\Gamma^*}_{k,r}|^2, \\
    &\leq 2 \sum_{k: v^*_k\in \mathcal{N}^{\mathcal{V}^*}_{k_r, r} \cup \{k_r\}} |c^{\Gamma^*}_{k,r}|^2. \nonumber
\end{align}
by the Cauchy-Schwarz inequality. With update filters satisfying $0 < b^{\Gamma^*}_{s,r} \leq \frac{1}{2}$, for all $v^*_s\in \mathcal{N}^{\mathcal{V}^*}_{k_r, r}$ \citep{jansen2009multiscale}, this guarantees that $c^{\Gamma^*}_{s,r-1}$ is bounded after the update (equation (\ref{equ:LG-update})) for all $v^*_s\in \mathcal{N}^{\mathcal{V}^*}_{k_r, r}$. For the one-level undo lifting, the boundness can be obtained by duality, see \cite{jansen2009multiscale}. From equation (\ref{equ:predict_bound}), note that the upper bound is given by the value $(1 + \sum_{s: v^*_s\in \mathcal{N}^{\mathcal{V}^*}_{k_r, r}} |a^{\Gamma^*}_{s,r}|^2)$. Hence, if we lift a new vertex which only has one neighbouring new vertex, the bound tends to be the maximum (2). This observation matches the practical sensitive points phenomena observed by \cite{jansen1998smoothing} and \cite{mahadevan2010multiscale}, where the term `sensitive points' indicates that boundary points contain high energy and have a significant influence on recovering the signal. On the other hand, if a new vertex with a large neighbourhood has been lifted and the prediction weights are almost evenly distributed (e.g. moving average), then the upper bound tends to be small.

\subsection{Scale Interpretation}
The notion of scale is essential for any wavelet-based method and relates to the level of detail, also known as the resolution, while also being strongly connected with the Fourier-based frequency \citep{nason2008wavelet, vidakovic2009statistical}.
While in the classical wavelet methods the scales can be generated by the dilation relation, within the LOCAAT framework the concept of scale is not directly discernible due to the absence of such relation. 
Inspired by the classical scale concept, the quantities in equation (\ref{equ:LG-decay}) can help in establishing a `scale'. Note that its expression is very similar to the `sum of distances' and the `average distance integral', thus, we simply define the scale of the detail coefficient $d^{\Gamma^*}_{k_r}$ obtained at stage-$r$ as 
\begin{align}
\label{equ:scale-choice}
    \text{scale}^{\Gamma^*}_{k_r} = I^{\Gamma^*}_{k_r, r}
\end{align}
The integral satisfies that $I^{\Gamma^*}_{k_{r-1}, r-1} \geq I^{\Gamma*}_{k_r, r}$, which implies the correspondence between removal order and scale. Note that when using the integral values as a sequence of ones, the potential advantage of this integral choice is that for the initial lifting recursions, the next stage-$(r-1)$ removal choice $v^*_{k_{r-1}}$ cannot be a neighbouring vertex of $v^*_{k_{r}}$. For example, once we remove $v^*_{k_m}$, after the integral update, the integral values of its neighbourhood will exceed the value one, which guarantees that another vertex (not in the neighbourhood) will be picked for removal, hence the network space is explored more efficiently.

\subsection{Original Domain Transformation}
\label{sec:inverse-LG}

While the overall LG-LOCAAT transform and the line graph mapping (from $G$ to $G^*$) are invertible, the existence of a line graph inverse cannot be guaranteed at every stage $r$. This is since while any graph has its unique corresponding line graph, not all graphs are line graphs \citep{bondy1976graph}.

\begin{figure}
    \centering
    \includegraphics[scale=0.66]{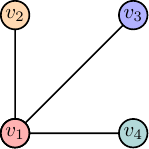}
    \caption{An example of a claw graph.}
    \label{fig:claw}
\end{figure}

As an example, let us consider the case where we lifted a new vertex $v^*_{k_r}$ at stage-$r$, and its neighbourhood is denoted as $\mathcal{N}^{\mathcal{V}^*}_{k_r, r}$. Therefore, we have the subgraph $G^{*\text{supp}}_{r} = (\mathcal{V}^{*\text{supp}}_{r}, \mathcal{E}^{*\text{supp}}_{r})$, where $\mathcal{V}^{*\text{supp}}_{r} = \{v^*_{k_r}\} \cup \mathcal{N}^{\mathcal{V}^*}_{k_r, r}$ and $\mathcal{E}^{*\text{supp}}_{r} = \{\{v^*_{k_r}, v^*_s\}\}_{s: \, v^*_s \in \mathcal{N}^{\mathcal{V}^*}_{k_r, r}}$. Recall that the prediction step at stage-$r$ is performed on this subgraph $G^{*\text{supp}}_{r}$, hence the analytic form of the wavelet function $\varphi^{\Gamma^*}_{k_r}$ is defined on the topology of $G^{*\text{supp}}_{r}$. Now if we want to obtain the associated analytical form defined on the original graph $G$ for $\varphi^{\Gamma^*}_{k_r}$, we have to find the inverse of $G^{*\text{supp}}_{r}$. Let us consider a special case, suppose there are three components in the neighbourhood $\mathcal{N}^{\mathcal{V}^*}_{k_r, r}$. This graph topology is referred to as `claw graph' in graph theory literature, see \cite{bondy1976graph} and Figure \ref{fig:claw}. 
Unfortunately, claw graphs are not line graphs \citep{chudnovsky2005structure}, which indicates that $G^{*\text{supp}}_{r}$ has no interpretation in the original graph domain if it is a claw graph. Moreover, any star graph with more than four vertices is not a line graph \citep{bondy1976graph}. Hence, at any stage-$r$, if there are more than three vertices used for prediction, then the associated topology $G^{*\text{supp}}_{r}$ will have no interpretation. This indicates that we cannot guarantee a graph correspondence between the original and line graph domains at each stage of the algorithm.

\section{Proofs}\label{app:proofs}

\subsection*{Proof for Proposition \ref{pro:LOCAAT-integral}}
\label{proof:proposition-LOCAAT-integral}
Starting with $C\cdot \underline{I}^* = \{C I^*_{k,m}\}_{k\in \{1,...m\}}$ will not change the stage-$m$ split step as long as $C > 0$, and the same new vertex $v^*_{k_m}$ will be lifted at this stage. Since the predict step is independent of the integral values, the value of the detail coefficient $d^*_{k_m}$ and the prediction weights remain the same. For any $s\in \mathcal{N}^{\mathcal{V}^*}_{k_m, m}$, the update of the integral sequence is now
\begin{align*}
    &C I^*_{s,m} + a^*_{s,m}  C I^*_{k_m,m}  \\
    = &C(I^*_{s,m} + a^*_{s,m} I^*_{k_m,m})  \\
    = &C I^*_{s,m-1},
\end{align*}
which indicates that in the stage-$(m-1)$, the integral sequence is proportional to the integral with the same multiplier constant $C$.
The $s$-th update coefficient is determined by the minimum norm solution such that
\begin{align*}
    b^{\Gamma^*}_{s, m} &= \frac{CI^*_{s,m-1} CI^*_{k_m, m}}{\sum_{t: v^*_t \in \mathcal{N}^{\mathcal{V}^*}_{k_m, m}} (CI^*_{t,m-1})^2}  \\
    &= \frac{I^*_{s,m-1} I^*_{k_m, m}}{\sum_{t: v^*_t \in \mathcal{N}^{\mathcal{V}^*}_{k_m, m}} (I^*_{t,m-1})^2},
\end{align*}
which coincides with the result when starting with $\underline{I}^*$. Thus, repeating the procedures above and by induction, we conclude that the detail coefficients and the filters do not change upon a proportional change in the integral values.

\section*{Proof for Proposition \ref{pro:LG-decay}}
\label{proof:proposition-LG-LOCAAT-decay}
Recall the detail coefficient is obtained by the prediction step, in which we have
\begin{align}
\label{equ:abs_detail}
    \left|d^{\Gamma^*}_{k_r}\right| = \left| c^{\Gamma^*}_{k_r,r}- \sum_{s: v^*_s \in \mathcal{N}^{\mathcal{V}^*}_{k_r, r}} a^{\Gamma^*}_{s,r} c^{\Gamma^*}_{s, r} \right|.
\end{align}
If for all $t:\, v^*_t \in \mathcal{N}^{\mathcal{V}^*}_{k_r, r}\cup \{v^*_{k_r}\}$, we have that $c^{\Gamma^*}_{t, r} = g^{\Gamma^*}_{t}$, then equation (\ref{equ:abs_detail}) can be written as
\begin{align}
    \left|d^{\Gamma^*}_{k_r}\right| &= \left| g^{\Gamma^*}_{k_r}- \sum_{s: v^*_s \in \mathcal{N}^{\mathcal{V}^*}_{k_r, r}} a^{\Gamma^*}_{s,r} g^{\Gamma^*}_{s} \right| \nonumber \\
    &= \left| \sum_{s: v^*_s \in \mathcal{N}^{\mathcal{V}^*}_{k_r, r}} a^{\Gamma^*}_{s,r} (g^{\Gamma^*}_{k_r}- g^{\Gamma^*}_{s}) \right|  \nonumber \\
    &\leq \sum_{s: v^*_s \in \mathcal{N}^{\mathcal{V}^*}_{k_r, r}} a^{\Gamma^*}_{s,r} \left| g^{\Gamma^*}_{k_r}- g^{\Gamma^*}_{s} \right|,
\end{align}
since $a^{\Gamma^*}_{s,r}\geq 0$ for all $s: v^*_s \in \mathcal{N}^{\mathcal{V}^*}_{k_r, r}$. Then because the function $g^{\Gamma^*}$ is Lipschitz with a constant $0< C < \infty $, then we have 
\begin{align}
    \left|d^{\Gamma^*}_{k_r}\right| &\leq \sum_{s: v^*_s \in \mathcal{N}^{\mathcal{V}^*}_{k_r, r}} a^{\Gamma^*}_{s,r} \left| g^{\Gamma^*}_{k_r}- g^{\Gamma^*}_{s} \right|  \nonumber \\
    \label{detail_inequality}
    &\leq \sum_{s: v^*_s \in \mathcal{N}^{\mathcal{V}^*}_{k_r, r}} a^{\Gamma^*}_{s,r}\, C \, \text{dist}(\mathbf{p}^*_{v^*_{k_r}}, \mathbf{p}^*_{v^*_s}). 
\end{align}
Recall that the prediction weights are normalised inverse distances, $a^{\Gamma^*}_{s,r} = \frac{1/\text{dist}(\mathbf{p}^*_{v^*_{k_r}}, \mathbf{p}^*_{v^*_s})}{\sum_{t: v^*_t \in \mathcal{N}^{\mathcal{V}^*}_{k_r, r}} 1/\text{dist}(\mathbf{p}^*_{v^*_{k_r}}, \mathbf{p}^*_{v^*_t})}$. Plug it into the inequality (\ref{detail_inequality}), we have
\begin{align}
    \left|d^{\Gamma^*}_{k_r}\right| &\leq \sum_{s: v^*_s \in \mathcal{N}^{\mathcal{V}^*}_{k_r, r}} \frac{1/\text{dist}(\mathbf{p}^*_{v^*_{k_r}}, \mathbf{p}^*_{v^*_s})}{\sum_{t: v^*_t \in \mathcal{N}^{\mathcal{V}^*}_{k_r, r}} 1/\text{dist}(\mathbf{p}^*_{v^*_{k_r}}, \mathbf{p}^*_{v^*_t})}\, C \, \text{dist}(\mathbf{p}^*_{v^*_{k_r}}, \mathbf{p}^*_{v^*_s})  \nonumber \\
    &= \sum_{s: v^*_s \in \mathcal{N}^{\mathcal{V}^*}_{k_r, r}} \frac{1}{\sum_{t: v^*_t \in \mathcal{N}^{\mathcal{V}^*}_{k_r, r}} 1/\text{dist}(\mathbf{p}^*_{v^*_{k_r}}, \mathbf{p}^*_{v^*_t})}\, C   \nonumber \\
    &= \frac{1}{|\mathcal{N}^{\mathcal{V}^*}_{k_r, r}|} \sum_{s: v^*_s \in \mathcal{N}^{\mathcal{V}^*}_{k_r, r}} \frac{|\mathcal{N}^{\mathcal{V}^*}_{k_r, r}|}{\sum_{t: v^*_t \in \mathcal{N}^{\mathcal{V}^*}_{k_r, r}} 1/\text{dist}(\mathbf{p}^*_{v^*_{k_r}}, \mathbf{p}^*_{v^*_t})}\, C   \nonumber \\
    &\leq \frac{1}{|\mathcal{N}^{\mathcal{V}^*}_{k_r, r}|} \sum_{s: v^*_s \in \mathcal{N}^{\mathcal{V}^*}_{k_r, r}} \frac{\sum_{t: v^*_t \in \mathcal{N}^{\mathcal{V}^*}_{k_r, r}} \text{dist}(\mathbf{p}^*_{v^*_{k_r}}, \mathbf{p}^*_{v^*_t})}{|\mathcal{N}^{\mathcal{V}^*}_{k_r, r}|}\, C \nonumber \\
    &= \frac{\sum_{t: v^*_t \in \mathcal{N}^{\mathcal{V}^*}_{k_r, r}} \text{dist}(\mathbf{p}^*_{v^*_{k_r}}, \mathbf{p}^*_{v^*_t})}{|\mathcal{N}^{\mathcal{V}^*}_{k_r, r}|}\, C,
\end{align}
since for positive $x_1, ..., x_n$, we have
\begin{align}
    \frac{n}{1/x_1 + ... + 1/x_n} \leq \frac{x_1 + ... + x_n}{n}.
\end{align}

\section{Supporting evidence for Section \ref{subsec:simresults}}\label{app:simresults}

\subsection{Stability Results}
\begin{table*}[hbt!]
    \centering
    \begin{tabular}{|c|c|c|c|c|c|}
    \hline
    Condition Number & Max & 75\% & Median & 25\% & Min \\
    \hline
    LG-Sid-c & 14.5314 & 12.9183 & 12.4962 & 11.9702 & 11.1918  \\
    \hline
    LG-Aid-c & 15.0632 & 13.2504 & 12.5252 & 11.9598 & 11.1996  \\
    \hline
    LG-Did-c & 13.9051 & 12.5774 & 11.5010 & 11.0700 & 10.6420  \\
    \hline
    LG-Snw-c & 13.5717 & 12.3798 & 11.7343 & 11.3743 & 10.8643  \\
    \hline
    LG-Anw-c & 12.7559 & 11.4412 & 11.0405 & 10.5475 & 10.0281  \\
    \hline
    \textbf{LG-Dnw-c} & \textbf{12.1607} & \textbf{11.0283} & \textbf{10.5684} & \textbf{10.2285} & \textbf{9.9500}  \\
    \hline
    \end{tabular}
    \vspace{0.3cm}
    \caption{Condition number for LG-LOCAAT with coordinate information.}
    \label{tab:LG_condition_number}
\end{table*}

\begin{table*}[hbt!]
    \centering
    \begin{tabular}{|c|c|c|c|c|c|}
    \hline
    Condition Number & Max & 75\% & Median & 25\% & Min \\
    \hline
    LG-Sid-p & 13.4308 & 12.3485 & 11.7877 & 11.3759 & 10.7340  \\
    \hline
    LG-Aid-p & 12.7611 & 11.4225 & 10.9914 & 10.5397 & 10.0863  \\
    \hline
    LG-Did-p & 12.6918 & 11.6651 & 10.7789 & 10.3077 & 10.0251  \\
    \hline
    LG-Snw-p & 13.2506 & 12.0824 & 11.5843 & 11.2258 & 10.6871  \\
    \hline
    LG-Anw-p & 12.5608 & 11.3372 & 10.7768 & 10.3567 & 10.0530  \\
    \hline
    \textbf{LG-Dnw-p} & \textbf{12.2235} & \textbf{11.0813} & \textbf{10.5528} & \textbf{10.1628} & \textbf{9.9535}  \\
    \hline
    \end{tabular}
    \vspace{0.3cm}
    \caption{Condition number for LG-LOCAAT using the path length.}
    \label{tab:LG_condition_number_without}
\end{table*}

\subsection{Sparsity Plots}

\begin{figure}[!h]
    \centering
    \includegraphics[width=8cm, height=12cm]{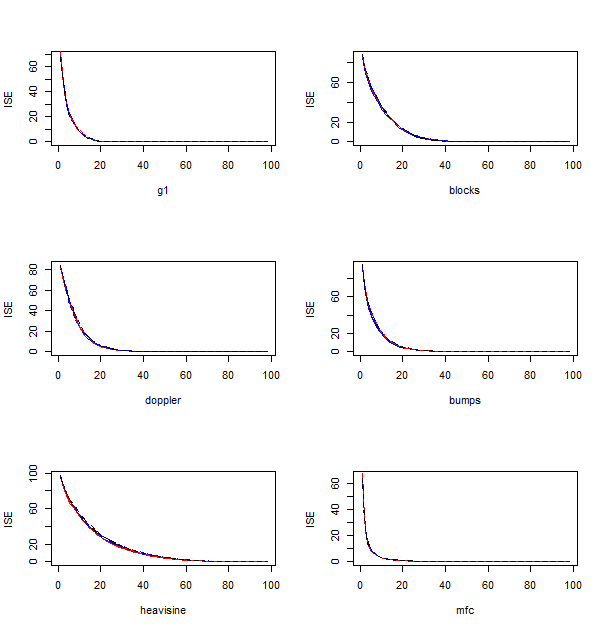}
    \caption{Sparsity plots for the test functions in equation (\ref{equ:test_func_normal}). The scheme is based on coordinate information. From left to right on {\em top row}: $g_1$, Blocks; {\em middle row}: Doppler, Bumps; {\em bottom row}: Heavisine, maartenfunc. \textbf{Black line}: LG-Sid-c; \textbf{red line}: LG-Aid-c; \textbf{blue line}: LG-Did-c; \textbf{dashed black line}: LG-Snw-c; \textbf{dashed red line}: LG-Anw-c; \textbf{dashed blue line}: LG-Dnw-c.}
    \label{fig:sparsity_coord_normal_func}
\end{figure}

\begin{figure}[!h]
    \centering
    \includegraphics[width=8cm, height=12cm]{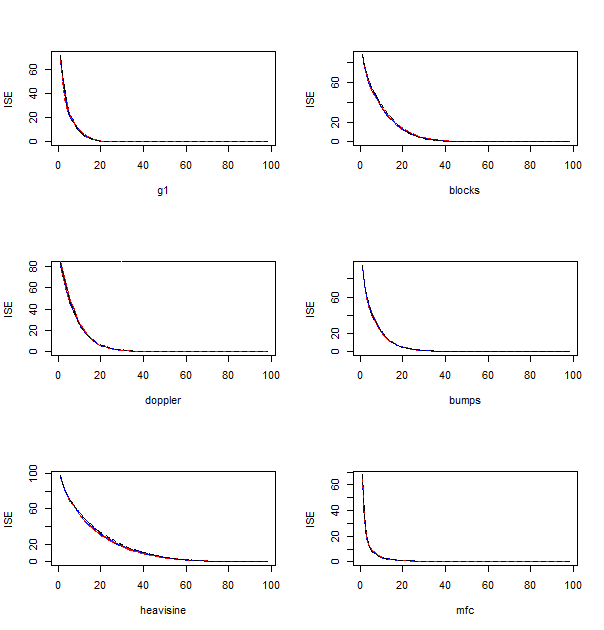}
    \caption{Sparsity plots for the test functions in equation (\ref{equ:test_func_normal}). The scheme is based on path distance. From left to right on {\em top row}: $g_1$, Blocks; {\em middle row}: Doppler, Bumps; {\em bottom row}: Heavisine, maartenfunc. \textbf{Black line}: LG-Sid-p; \textbf{red line}: LG-Aid-p; \textbf{blue line}: LG-Did-p; \textbf{dashed black line}: LG-Snw-p; \textbf{dashed red line}: LG-Anw-p; \textbf{dashed blue line}: LG-Dnw-p.}
    \label{fig:sparsity_path_normal_func}
\end{figure}

\begin{figure}[!h]
    \centering
    \includegraphics[width=8cm, height=12cm]{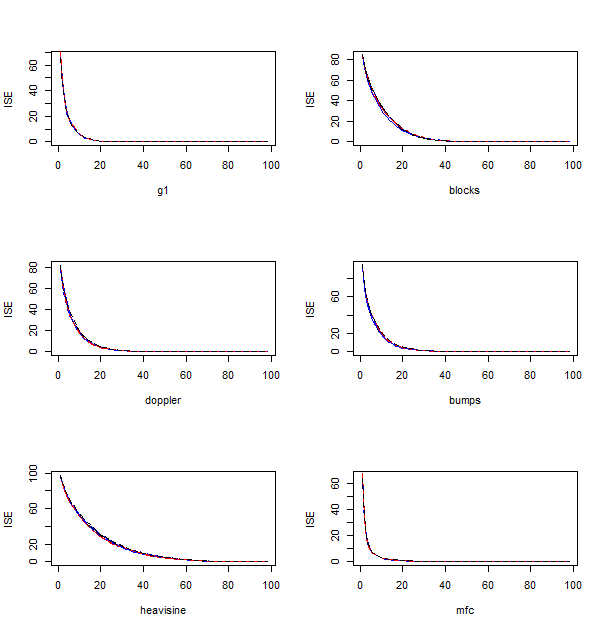}
    \caption{Sparsity plots for the test functions in equation (\ref{equ:test_func_edge_average}). The scheme is based on coordinate information. From left to right on {\em top row}: $g_1$, Blocks; {\em middle row}: Doppler, Bumps; {\em bottom row}: Heavisine, maartenfunc. \textbf{Black line}: LG-Sid-c; \textbf{red line}: LG-Aid-c; \textbf{blue line}: LG-Did-c; \textbf{dashed black line}: LG-Snw-c; \textbf{dashed red line}: LG-Anw-c; \textbf{dashed blue line}: LG-Dnw-c.}
    \label{fig:sparsity_coord_edge_average}
\end{figure}

\begin{figure}[!h]
    \centering
    \includegraphics[width=8cm, height=12cm]{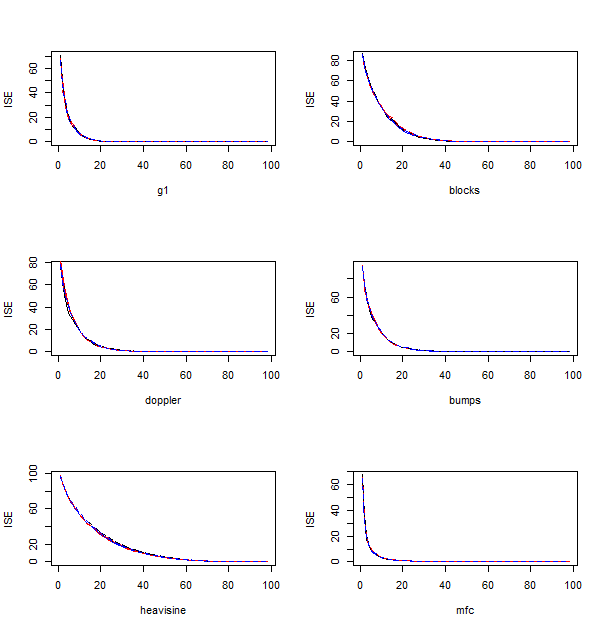}
    \caption{Sparsity plots for the test functions in equation (\ref{equ:test_func_edge_average}). The scheme is based on path distance. From left to right on {\em top row}: $g_1$, Blocks; {\em middle row}: Doppler, Bumps; {\em bottom row}: Heavisine, maartenfunc. \textbf{Black line}: LG-Sid-p; \textbf{red line}: LG-Aid-p; \textbf{blue line}: LG-Did-p; \textbf{dashed black line}: LG-Snw-p; \textbf{dashed red line}: LG-Anw-p; \textbf{dashed blue line}: LG-Dnw-p.}
    \label{fig:sparsity_path_edge_average}
\end{figure}

\subsection{Denoising Performance Tables}


\begin{table*}[hbt!]
    \centering
    \scalebox{1}{
    \begin{tabular}{|c|c|c|c|c|c|c|c|c|}
    \hline
    Variance$\times 10^3$ & $g_1$ & Blocks & Doppler & Bumps & Heavisine & mfc \\
    \hline
    \multicolumn{6}{c}{SNR=3}    \\
    \hline
    LG-Sid-c & 48 & 65 & 61 & 51 & 68 & 39 \\
    \hline
    \textbf{LG-Aid-c} & \textbf{46} & \textbf{64} & \textbf{55} & \textbf{48} & \textbf{64} & \textbf{36} \\
    \hline
    LG-Did-c & 49 & 81 & 66 & 56 & 101 & 37 \\
    \hline
    LG-Snw-c & 49 & 66 & 61 & 52 & 69 & 39 \\
    \hline
    \textbf{LG-Anw-c} & \textbf{46} & \textbf{64} & \textbf{55} & \textbf{48} & \textbf{64} & \textbf{36} \\
    \hline
    LG-Dnw-c & 52 & 86 & 71 & 60 & 113 & 39 \\
    \hline
    \multicolumn{6}{c}{SNR=5}    \\
    \hline
    LG-Sid-c & \textbf{18} & \textbf{24} & 23 & 22 & 31 & 15 \\
    \hline
    LG-Aid-c & \textbf{18} & \textbf{24} & 22 & \textbf{20} & 27 & \textbf{14} \\
    \hline
    LG-Did-c & 20 & 33 & 27 & 26 & 67 & \textbf{15} \\
    \hline
    LG-Snw-c & 19 & 25 & 23 & 22 & 31 & 15 \\
    \hline
    \textbf{LG-Anw-c} & \textbf{18} & \textbf{24} & \textbf{21} & \textbf{20} & \textbf{26} & \textbf{14} \\
    \hline
    LG-Dnw-c & 21 & 34 & 30 & 28 & 78 & 17 \\
    \hline
    \multicolumn{6}{c}{SNR=7}    \\
    \hline
    LG-Sid-c & \textbf{9 }& \textbf{13} & \textbf{12} & 12 & 18 & 9 \\
    \hline
    \textbf{LG-Aid-c} & \textbf{9 }& \textbf{13} & \textbf{12} & \textbf{11} & \textbf{15} & \textbf{8 }\\
    \hline
    LG-Did-c & 10 & 18 & 15 & 16 & 57 & 9 \\
    \hline
    LG-Snw-c & \textbf{9 }& \textbf{13} & \textbf{12} & 12 & 18 & 9 \\
    \hline
    LG-Anw-c & \textbf{9 }& \textbf{13} & \textbf{12} & 12 & \textbf{15} & \textbf{8 }\\
    \hline
    LG-Dnw-c & 10 & 18 & 16 & 17 & 68 & 10 \\
    \hline    
    \end{tabular}}
    \vspace{0.3cm}
    \caption{Variance for LG-LOCAAT on a tree structure with 100 nodes and 99 edges. The functions follow the pointwise construction. We assume the coordinate information is available. }
    \label{tab:edge_LG_variance_99-coordinate}
\end{table*}

\begin{table*}[hbt!]
    \centering
    \scalebox{1}{
    \begin{tabular}{|c|c|c|c|c|c|c|c|c|}
    \hline
    $\text{Bias}^2 \times 10^3$ & $g_1$ & Blocks & Doppler & Bumps & Heavisine &mfc \\
    \hline
    \multicolumn{6}{c}{SNR=3}    \\
    \hline
    LG-Sid-c & 17 & 49 & 48 & 41 & 204 & 13 \\
    \hline
    LG-Aid-c & 17 & 49 & 37 & \textbf{32} & \textbf{134} & 9\\
    \hline
    \textbf{LG-Did-c} & \textbf{14} & \textbf{39} & \textbf{27} & 24 & \textbf{134} & \textbf{8} \\
    \hline
    LG-Snw-c & 17 & 51 & 48 & 44 & 213 & 14 \\
    \hline
    LG-Anw-c & 18 & 51 & 37 & 33 & 141 & 10 \\
    \hline
    LG-Dnw-c & 15 & 40 & \textbf{27} & \textbf{23} & 148 & \textbf{8} \\
    \hline
    \multicolumn{6}{c}{SNR=5}    \\
    \hline
    LG-Sid-c & 4& 17 & 20 & 23 & 175 & 10 \\
    \hline
    LG-Aid-c & 5 & 18 & 16 & 17 & 112 & 7 \\
    \hline
    \textbf{LG-Did-c} & \textbf{3} & \textbf{13} & \textbf{10} & \textbf{13} & \textbf{111} & \textbf{6}\\
    \hline
    LG-Snw-c & 4& 17 & 21 & 25 & 186 & 11 \\
    \hline
    LG-Anw-c & 5 & 19 & 17 & 18 & 121 & 8 \\
    \hline
    LG-Dnw-c & \textbf{3} & \textbf{13} & \textbf{10} & \textbf{13} & 129 & \textbf{6} \\
    \hline
    \multicolumn{6}{c}{SNR=7}    \\
    \hline
    LG-Sid-c & 2 & 9 & 12 & 16 & 166 & 9 \\
    \hline
    LG-Aid-c & 2 & 9 & 9 & 11 & 105 & \textbf{6} \\
    \hline
    \textbf{LG-Did-c} & \textbf{1} & \textbf{7} & \textbf{5} & \textbf{8}& \textbf{104} & \textbf{5} \\
    \hline
    LG-Snw-c & \textbf{1} & 9 & 12 & 16 & 177 & 9 \\
    \hline
    LG-Anw-c & 2 & 10 & 10 & 12 & 115 & 6 \\
    \hline
    LG-Dnw-c & \textbf{1} & \textbf{7} & \textbf{5} &\textbf{8} & 124 & \textbf{5} \\
    \hline
    \end{tabular}}
    \vspace{0.3cm}
    \caption{Squared bias for LG-LOCAAT on a tree structure with 100 nodes and 99 edges. The functions follow the pointwise construction. We assume the coordinate information is available. }
    \label{tab:edge_LG_bias_99-coordinate}
\end{table*}


\begin{table*}[hbt!]
    \centering
    \scalebox{1}{
    \begin{tabular}{|c|c|c|c|c|c|c|c|c|}
    \hline
    Variance$\times 10^3$ & $g_1$ & Blocks & Doppler & Bumps & Heavisine & mfc \\
    \hline
    \multicolumn{6}{c}{SNR=3}    \\
    \hline
    LG-Sid-p & 49 & 66 & 61 & 52 & 69 & 39 \\
    \hline
    \textbf{LG-Aid-p} & \textbf{47} & \textbf{64} & 57 & \textbf{49} & \textbf{63} & \textbf{36} \\
    \hline
    LG-Did-p & 51 & 83 & 69 & 60 & 113 & 39 \\
    \hline
    LG-Snw-p & 49 & 67 & 61 & 52 & 70 & 39 \\
    \hline
    LG-Anw-p & \textbf{47} & 65 & \textbf{56} & 50 & 64 & 37 \\
    \hline
    LG-Dnw-p & 53 & 88 & 73 & 63 & 121 & 40 \\
    \hline
    \multicolumn{6}{c}{SNR=5}    \\
    \hline
    LG-Sid-p & 19 & 25 & 23 & 22 & 31 & 16 \\
    \hline
    \textbf{LG-Aid-p} & \textbf{18} & \textbf{24} & \textbf{22} & \textbf{21} & \textbf{26} & \textbf{15} \\
    \hline
    LG-Did-p & 21 & 34 & 29 & 28 & 77 & 17 \\
    \hline
    LG-Snw-p & 19 & 25 & 23 & 22 & 31 & 16 \\
    \hline
    LG-Anw-p & \textbf{18} & 25 & \textbf{22} & \textbf{21} & \textbf{26} & \textbf{15} \\
    \hline
    LG-Dnw-p & 21 & 35 & 30 & 29 & 85 & 18 \\
    \hline
    \multicolumn{6}{c}{SNR=7}    \\
    \hline
    LG-Sid-p & \textbf{9} & \textbf{13} & 12 & \textbf{12} & 18 & 9 \\
    \hline
    \textbf{LG-Aid-p} & \textbf{9} & \textbf{13} & \textbf{11} & \textbf{12} & \textbf{15} & \textbf{8}\\
    \hline
    LG-Did-p & 10 & 18 & 16 & 17 & 65 & 10 \\
    \hline
    LG-Snw-p & \textbf{9} & \textbf{13} & 12 & 13 & 19 & 9 \\
    \hline
    LG-Anw-p & \textbf{9} & \textbf{13} & 12 & \textbf{12} & \textbf{15} & \textbf{8}\\
    \hline
    LG-Dnw-p & 10 & 19 & 17 & 18 & 74 & 11 \\
    \hline
    \end{tabular}}
    \vspace{0.3cm}
    \caption{Variance for LG-LOCAAT on a tree structure with 100 nodes and 99 edges. The functions follow the pointwise construction. The path distance is used.}
    \label{tab:edge_LG_variance_99-path}
\end{table*}

\begin{table*}[hbt!]
    \centering
    \scalebox{1}{
    \begin{tabular}{|c|c|c|c|c|c|c|c|c|}
    \hline
    $\text{Bias}^2$ & $g_1$ & Blocks & Doppler & Bumps & Heavisine &mfc \\
    \hline
    \multicolumn{6}{c}{SNR=3}    \\
    \hline
    LG-Sid-p & 17 & 50 & 48 & 42 & 205 & 14 \\
    \hline
    LG-Aid-p & 18 & 53 & 37 & 35 & 145 & 11 \\
    \hline
    \textbf{LG-Did-p} & \textbf{14} & \textbf{39} & \textbf{26} & \textbf{23} & \textbf{140} & \textbf{8} \\
    \hline
    LG-Snw-p & 19 & 51 & 49 & 43 & 218 & 15 \\
    \hline
    LG-Anw-p & 19 & 58 & 39 & 35 & 151 & 11 \\
    \hline
    LG-Dnw-p & 15 & 41 & 27 & \textbf{23} & 155 & \textbf{8} \\
    \hline
    \multicolumn{6}{c}{SNR=5}    \\
    \hline
    LG-Sid-p & 4 & 17 & 21 & 24 & 177 & 11 \\
    \hline
    LG-Aid-p & 5 & 20 & 16 & 19 & 126 & 8 \\
    \hline
    \textbf{LG-Did-p} & \textbf{3} & \textbf{13} & \textbf{9} & \textbf{12} & \textbf{115} & \textbf{6} \\
    \hline
    LG-Snw-p & 4 & 17 & 21 & 25 & 188 & 12 \\
    \hline
    LG-Anw-p & 5 & 21 & 17 & 19 & 132 & 9 \\
    \hline
    LG-Dnw-p & \textbf{3} & \textbf{13} & \textbf{9} & \textbf{12} & 133 & \textbf{6} \\
    \hline
    \multicolumn{6}{c}{SNR=7}    \\
    \hline
    LG-Sid-p & 2 & 9 & 12 & 16 & 168 & 9 \\
    \hline
    LG-Aid-p & 2 & 10 & 10 & 12 & 120 & 6 \\
    \hline
    \textbf{LG-Did-p} & \textbf{1} & \textbf{7} & \textbf{5} & \textbf{8} & \textbf{107} & \textbf{5} \\
    \hline
    LG-Snw-p & 2 & 9 & 12 & 16 & 178 & 9 \\
    \hline
    LG-Anw-p & 2 & 11 & 10 & 13 & 127 & 7 \\
    \hline
    LG-Dnw-p & \textbf{1} & \textbf{7} & \textbf{5} & \textbf{8} & 127 & \textbf{5} \\
    \hline
    \end{tabular}}
    \vspace{0.3cm}
    \caption{Squared bias for LG-LOCAAT on a tree structure with 100 nodes and 99 edges. The functions follow the pointwise construction. The path distance is used. }
    \label{tab:edge_LG_bias_99-path}
\end{table*}


\begin{table*}[hbt!]
    \centering
    \scalebox{1}{
    \begin{tabular}{|c|c|c|c|c|c|c|c|c|}
    \hline
    Variance$\times 10^3$ & $g_1$ & Blocks & Doppler & Bumps & Heavisine & mfc \\
    \hline
    \multicolumn{6}{c}{SNR=3}    \\
    \hline
    LG-Sid-c & 45 & 65 & 56 & 51 & 67 & 38 \\
    \hline
    \textbf{LG-Aid-c} & \textbf{42} & \textbf{62} & 52 & \textbf{47} & \textbf{64} & \textbf{36} \\
    \hline
    LG-Did-c & 46 & 78 & 60 & 55 & 99 & 37 \\
    \hline
    LG-Snw-c & 46 & 65 & 56 & 52 & 67 & 38 \\
    \hline
    LG-Anw-c & \textbf{42} & 63 & \textbf{51} & 48 & \textbf{64} & \textbf{36} \\
    \hline
    LG-Dnw-c & 48 & 84 & 65 & 59 & 111 & 39 \\
    \hline
    \multicolumn{6}{c}{SNR=5}    \\
    \hline
    LG-Sid-c & 18 & 26 & 23 & 22 & 30 & 15 \\
    \hline
    \textbf{LG-Aid-c} & \textbf{17} & \textbf{25} & \textbf{21} & \textbf{20} & 27 & \textbf{14} \\
    \hline
    LG-Did-c & 20 & 35 & 27 & 26 & 65 & 15 \\
    \hline
    LG-Snw-c & 18 & 26 & 23 & 22 & 30 & 15 \\
    \hline
    LG-Anw-c & 18 & 26 & \textbf{21} & \textbf{20} & \textbf{26} & \textbf{14} \\
    \hline
    LG-Dnw-c & 21 & 37 & 29 & 28 & 77 & 17 \\
    \hline
    \multicolumn{6}{c}{SNR=7}    \\
    \hline
    LG-Sid-c & 10 & 14 & \textbf{12} & 12 & 18 & \text{8} \\
    \hline
    \textbf{LG-Aid-c} & \textbf{9} & \textbf{13} & \textbf{12} & \textbf{11} & \textbf{15} & \text{8} \\
    \hline
    LG-Did-c & 11 & 19 & 15 & 16 & 55 & 9 \\
    \hline
    LG-Snw-c & 10 & 14 & \textbf{12} & 12 & 17 & 9 \\
    \hline
    LG-Anw-c & \textbf{9} & \textbf{13} & \textbf{12} & \textbf{11} & \textbf{15} & \text{8} \\
    \hline
    LG-Dnw-c & 11 & 21 & 17 & 17 & 67 & 10 \\
    \hline
    \end{tabular}}
    \vspace{0.3cm}
    \caption{Variance for LG-LOCAAT on a tree structure with 100 nodes and 99 edges. The functions follow the edge-averaging construction. We assume the coordinate information is available. }
    \label{tab:edge_average_LG_variance_99-coordinate}
\end{table*}

\begin{table*}[hbt!]
    \centering
    \scalebox{1}{
    \begin{tabular}{|c|c|c|c|c|c|c|c|c|}
    \hline
    $\text{Bias}^2$ & $g_1$ & Blocks & Doppler & Bumps & Heavisine &mfc \\
    \hline
    \multicolumn{6}{c}{SNR=3}    \\
    \hline
    LG-Sid-c & 19 & 61 & 47 & 41 & 215 & 12 \\
    \hline
    LG-Aid-c & 17 & 58 & 37 & 31 & 142 & 9 \\
    \hline
    \textbf{LG-Did-c} & \textbf{13} & \textbf{42} & \textbf{28} & \textbf{23} & \textbf{130} & \textbf{7} \\
    \hline
    LG-Snw-c & 20 & 63 & 48 & 44 & 224 & 13 \\
    \hline
    LG-Anw-c & 18 & 62 & 39 & 32 & 152 & 9 \\
    \hline
    LG-Dnw-c & \textbf{13} & 44 & \textbf{28} & \textbf{23} & 144 & \textbf{7} \\
    \hline
    \multicolumn{6}{c}{SNR=5}    \\
    \hline
    LG-Sid-c & 7 & 25 & 22 & 24 & 190 & 10 \\
    \hline
    LG-Aid-c & 7 & 24 & 18 & 18 & 120 & 7 \\
    \hline
    \textbf{LG-Did-c} & \textbf{5} & \textbf{17} & \textbf{11} & \textbf{12} & \textbf{111} & \textbf{6} \\
    \hline
    LG-Snw-c & 7 & 25 & 23 & 25 & 201 & 11 \\
    \hline
    LG-Anw-c & 8 & 25 & 19 & 19 & 131 & 7 \\
    \hline
    LG-Dnw-c & \textbf{5} & \textbf{17} & \textbf{11} & \textbf{12} & 128 & \textbf{6} \\
    \hline
    \multicolumn{6}{c}{SNR=7}    \\
    \hline
    LG-Sid-c & 3 & 13 & 14 & 16 & 181 & 9 \\
    \hline
    LG-Aid-c & 3 & 13 & 11 & 12 & 114 & 6 \\
    \hline
    \textbf{LG-Did-c} & \textbf{2} & \textbf{9} & \textbf{6} & \textbf{8} & \textbf{105} & \textbf{4} \\
    \hline
    LG-Snw-c & 3 & 14 & 14 & 17 & 193 & 9 \\
    \hline
    LG-Anw-c & 4 & 14 & 11 & 12 & 126 & 6 \\
    \hline
    LG-Dnw-c & \textbf{2} & \textbf{9} & \textbf{6} & \textbf{8} & 123 & 5 \\
    \hline
    \end{tabular}}
    \vspace{0.3cm}
    \caption{Squared bias for LG-LOCAAT on a tree structure with 100 nodes and 99 edges. The functions follow the edge-averaging construction. We assume the coordinate information is available. }
    \label{tab:edge_average_LG_bias_99-coordinate}
\end{table*}


\begin{table*}[hbt!]
    \centering
    \scalebox{1}{
    \begin{tabular}{|c|c|c|c|c|c|c|c|c|}
    \hline
    Variance$\times 10^3$ & $g_1$ & Blocks & Doppler & Bumps & Heavisine & mfc \\
    \hline
    \multicolumn{6}{c}{SNR=3}    \\
    \hline
    LG-Sid-p & 46 & 65 & 56 & 52 & 67 & 39 \\
    \hline
    \textbf{LG-Aid-p} & \textbf{43} & \textbf{63} & 53 & \textbf{49} & \textbf{63} & \textbf{36} \\
    \hline
    LG-Did-p & 48 & 81 & 63 & 59 & 111 & 39 \\
    \hline
    LG-Snw-p & 46 & 65 & 56 & 52 & 69 & 39 \\
    \hline
    LG-Anw-p & \textbf{43} & 64 & \textbf{52} & \textbf{49} & 64 & \textbf{36} \\
    \hline
    LG-Dnw-p & 49 & 86 & 67 & 62 & 119 & 40 \\
    \hline
    \multicolumn{6}{c}{SNR=5}    \\
    \hline
    LG-Sid-p & 19 & 26 & 23 & 22 & 30 & \textbf{15} \\
    \hline
    \textbf{LG-Aid-p} & \textbf{18} & \textbf{25} & \textbf{21} & \textbf{21} & \textbf{26} & \textbf{15} \\
    \hline
    LG-Did-p & 20 & 36 & 29 & 28 & 76 & 17 \\
    \hline
    LG-Snw-p & 19 & 26 & 23 & 22 & 30 & 16 \\
    \hline
    LG-Anw-p & \textbf{18} & 26 & 22 & \textbf{21} & \textbf{26} & \textbf{15} \\
    \hline
    LG-Dnw-p & 21 & 38 & 30 & 29 & 84 & 18 \\
    \hline
    \multicolumn{6}{c}{SNR=7}    \\
    \hline
    LG-Sid-p & 10 & \textbf{14} & \textbf{12} & \textbf{12} & 18 & 9 \\
    \hline
    \textbf{LG-Aid-p} & \textbf{9} & \textbf{14} & \textbf{12} & \textbf{12} & \textbf{15} & \textbf{8} \\
    \hline
    LG-Did-p & 11 & 20 & 16 & 17 & 65 & 10 \\
    \hline
    LG-Snw-p & 10 & \textbf{14} & 13 & \textbf{12} & 18 & 9 \\
    \hline
    LG-Anw-p & 10 & \textbf{14} & \textbf{12} & \textbf{12} & \textbf{15} & \textbf{8} \\
    \hline
    LG-Dnw-p & 12 & 21 & 17 & 18 & 74 & 11 \\
    \hline
    \end{tabular}}
    \vspace{0.3cm}
    \caption{Variance for LG-LOCAAT on a tree structure with 100 nodes and 99 edges. The functions follow the edge-averaging construction. The path distance is used. }
    \label{tab:edge_average_LG_variance_99-path}
\end{table*}

\begin{table*}[hbt!]
    \centering
    \scalebox{1}{
    \begin{tabular}{|c|c|c|c|c|c|c|c|c|}
    \hline
    $\text{Bias}^2$ & $g_1$ & Blocks & Doppler & Bumps & Heavisine &mfc \\
    \hline
    \multicolumn{6}{c}{SNR=3}    \\
    \hline
    LG-Sid-p & 20 & 62 & 47 & 42 & 215 & 14 \\
    \hline
    LG-Aid-p & 18 & 61 & 38 & 34 & 155 & 10 \\
    \hline
    \textbf{LG-Did-p} & \textbf{13} & \textbf{41} & \textbf{27} & \textbf{23} & \textbf{136} & \textbf{7} \\
    \hline
    LG-Snw-p & 20 & 64 & 48 & 44 & 227 & 14 \\
    \hline
    LG-Anw-p & 20 & 69 & 40 & 35 & 162 & 10 \\
    \hline
    LG-Dnw-p & \textbf{13} & 44 & 28 & \textbf{23} & 151 & \textbf{7} \\
    \hline
    \multicolumn{6}{c}{SNR=5}    \\
    \hline
    LG-Sid-p & 8 & 25 & 22 & 24 & 192 & 11 \\
    \hline
    LG-Aid-p & 8 & 26 & 18 & 19 & 135 & 8 \\
    \hline
    \textbf{LG-Did-p} & \textbf{4} & \textbf{16} & \textbf{11} & \textbf{12} & \textbf{115} & \textbf{6} \\
    \hline
    LG-Snw-p & 8 & 26 & 23 & 25 & 203 & 11 \\
    \hline
    LG-Anw-p & 8 & 29 & 19 & 19 & 144 & 8 \\
    \hline
    LG-Dnw-p & 5 & 17 & \textbf{11} & \textbf{12} & 133 & \textbf{6} \\
    \hline
    \multicolumn{6}{c}{SNR=7}    \\
    \hline
    LG-Sid-p & 3 & 14 & 13 & 16 & 185 & 9 \\
    \hline
    LG-Aid-p & 4 & 15 & 11 & 12 & 130 & 6 \\
    \hline
    \textbf{LG-Did-p} & \textbf{2} & \textbf{9} & \textbf{6} & \textbf{7} & \textbf{108} & \textbf{4} \\
    \hline
    LG-Snw-p & 3 & 14 & 14 & 17 & 194 & 9 \\
    \hline
    LG-Anw-p & 4 & 15 & 12 & 13 & 138 & 6 \\
    \hline
    LG-Dnw-p & \textbf{2} & \textbf{9} & \textbf{6} & 8 & 127 & 5 \\
    \hline
    \end{tabular}}
    \vspace{0.3cm}
    \caption{Squared bias for LG-LOCAAT on a tree structure with 100 nodes and 99 edges. The functions follow the edge-averaging construction. The path distance is used.}
    \label{tab:edge_average_LG_bias_99-path}
\end{table*}

\end{appendices}

\end{document}